\RequirePackage{fix-cm}
\documentclass[twocolumn,final]{svjour3}
\smartqed  
\usepackage{graphicx}
\usepackage{caption}  
\usepackage{subcaption}   
\usepackage{color, colortbl}   
\definecolor{Gray}{gray}{0.9}    
\usepackage{wrapfig}        
\usepackage{authorbiography}
\usepackage{hyperref}
\usepackage{blindtext}
\usepackage{tikz}
\usepackage{adjustbox}  
\usepackage{picins} 
\usepackage{algorithm}
\usepackage{algpseudocode}

\usepackage{multirow}  
\usepackage{siunitx}   

\usepackage[figuresright]{rotating}   
\begin{document}

\title{A Systematic Review of EEG-based Machine Intelligence Algorithms for Depression Diagnosis, and Monitoring}

\titlerunning{}        

\author{Amir Nassibi\and
        Christos Papavassiliou\and
        Ildar Rakhmatulin\and
        Danilo Mandic\and
        S. Farokh Atashzar}

\institute{Amir Nassibi (corresponding author)  \at
              Department of Electrical and Electronic Engineering, Imperial College London, London, UK.
              \email{a.nassibi15@imperial.ac.uk}           
           \and
           Christos Papavassiliou \at
              Department of Electrical and Electronic Engineering, Imperial College London, London, UK.
              \email{c.papavas@imperial.ac.uk}
                          \and
            Ildar Rakhmatulin  \at
              School of Engineering, University of Edinburgh, UK.
              \email{ildarr2016@gmail.com }
            \and
            Danilo Mandic \at
              Department of Electrical and Electronic Engineering, Imperial College London, London, UK.
              \email{d.mandic@imperial.ac.uk}
              \and
           S. Farokh Atashzar \at
              Department of  Electrical and Computer Engineering, New York University (NYU),USA.
              \email{f.atashzar@nyu.edu} 
              }

\maketitle

\begin{abstract}
Depression disorder is a serious health condition that has affected the lives of millions of people around the world. Diagnosis of depression is a challenging practice that relies heavily on subjective studies and, in most cases, suffers from late findings. Electroencephalography (EEG) biomarkers have been suggested and investigated in recent years as a potential transformative objective practice. In this article, for the first time, a detailed systematic review of EEG-based depression diagnosis approaches is conducted using advanced machine learning techniques and statistical analyses. For this, 938 potentially relevant articles (since 1985) were initially detected and filtered into 139 relevant articles based on the review scheme “preferred reporting items for systematic reviews and meta-analyses (PRISMA).” This article compares and discusses the selected articles and categorizes them according to the type of machine learning techniques and statistical analyses. Algorithms, preprocessing techniques, extracted features, and data acquisition systems are discussed and summarized. This review paper explains the existing challenges of the current algorithms and sheds light on the future direction of the field. This systematic review outlines the issues and challenges in machine intelligence for the diagnosis of EEG depression that can be addressed in future studies and possibly in future wearable technologies.

\end{abstract}




\section{Introduction}

Depression disorders currently affect 264 million people worldwide, according to a systematic analysis by J. Spencer et al.\cite{James2018}. The Adult Psychiatric Morbidity Survey (APMS) \cite{McManusSBebbingtonPJenkinsR2016} in England reveals that 8 percent of the population is diagnosed with depression or anxiety. In the US, a survey by the Center for Behavioral Health Statistics and Quality \cite{CenterforBehavioralHealthStatisticsandQuality20182017Tables} indicates that almost 7.1 percent of the US population has depression disorders.
Clinical interviews and questionnaires are used to diagnose depression in clinics. Traditional practice, on the other hand, makes early detection of depression unattainable \cite{LarryCulpepperMD2014MisdiagnosisPractices}. Furthermore, existing procedures make it impossible to continuously monitor patients or investigate the neurophysiological indicators of depressed patients. 
In recent years, physiological biomarkers that can be collected from signals such as electroencephalography (EEG) have been introduced as a potential objective measure of both cognitive and emotional states \cite{10.1007/978-1-4020-8387-7_120}, \cite{doi:10.1142/S0129065716500052}. These biomarkers may also have the power to detect early signs of depression due to the corresponding cortical changes. In several studies, researchers have used supervised machine learning algorithms to discriminate depressed patients and healthy subjects (such as \cite{8101151}, \cite{TekinErguzel2015}, \cite{Mumtaz2015ADisorder}, \cite{Puk2016}); Furthermore, statistical analysis is also used in several research articles to show the correlation between some biomarkers of EEG and diagnosed depression without the implementation of machine learning modules (such as \cite{Bachmann2015}, \cite{12424821}, \cite{Liao2013}, \cite{Hosseinifard2013}). 
This line of research and development has been motivated by the successful implementation of EEG-based diagnostic techniques for several other neurological disorders that in many cases also manifest depression-related symptoms such as schizophrenia \cite{Raghavendra2010},\cite{Raghavendra2009},\cite{QinglinZhao2013}, Alzheimer's disease \cite{Kim2005},\cite{Tylova2013},\cite{Sole-Casals2015},\cite{JiangZheng-Yan2004}, Parkinson's disease \cite{Hirschauer2015},\cite{Yuvaraj2016}, and Mild Cognitive Impairment (MCI) \cite{OKeeffe2017},\cite{Gomez2018},\cite{WeiLing2014}. Unlike the aforementioned conditions and despite strong evidence, to date, there are a few systematic reviews available in machine intelligence for the EEG-based diagnosis of depression. 
Al-Nafjan et al. \cite{Al-Nafjan2017} conducted a systematic review on EEG-based emotion recognition using brain computer interface systems. The authors reviewed 285 articles and discuss several different neurological disorders, including schizophrenia, disorders of consciousness, depression, and autism. In another study A. Shatte et al. \cite{Shatte2019} conducted a review of the methods and applications of using machine learning in mental health diagnosis. The authors studied 300 articles based on the most common mental health conditions such as schizophrenia, depression, and Alzheimer’s disease. 
Although the authors in \cite{Al-Nafjan2017} and \cite{Shatte2019} proposed a systematic review on mental health conditions including depression, they do not provide details such as preprocessing techniques, performance metrics of each algorithm, duration of analysis, participants selection criteria and details, experiment condition, number of electrodes, sampling frequency, electrode types and the type of filters and linear and non-linear features extracted in each experiment. In this section, a detailed comprehensive systematic review of EEG depression diagnosis from 1985 till December 2022 has been conducted with a focus on machine intelligence algorithms and statistical analysis to distinguish depressed patients from healthy subjects based on EEG biomarkers. 

The paper is organized as follows. In Methodology, the procedure and criteria to choose and filter relevant articles based on 'preferred reporting items for systematic reviews and meta-analyses (PRISMA)” \cite{Moher2009} systematic review method are discussed. In the next Section, we provided a brief overview of the key machine learning methodologies, including Supervised, unsupervised, semi-supervised, and reinforcement learning. This is followed by a summary of Machine Learning Algorithms such as Support Vector Machine (SVM), K-Nearest Neighbors (KNN), Decision Trees, Random Forests, Logistic regression and Neural networks (NN). Then a detailed and comprehensive analysis and review of papers based on sensors, machine learning algorithms, and statistical analysis are introduced. In this section, relevant papers are categorized and summarized in four tables, allowing a detailed comparative analysis of the existing work provided for the first time on this topic. Finally, in the last section, discussions are provided, the review is summarized, and the challenges and future vision are analyzed. 

\subsection{Methodology}

\begin{figure*}
\centering
\includegraphics[scale=0.7]{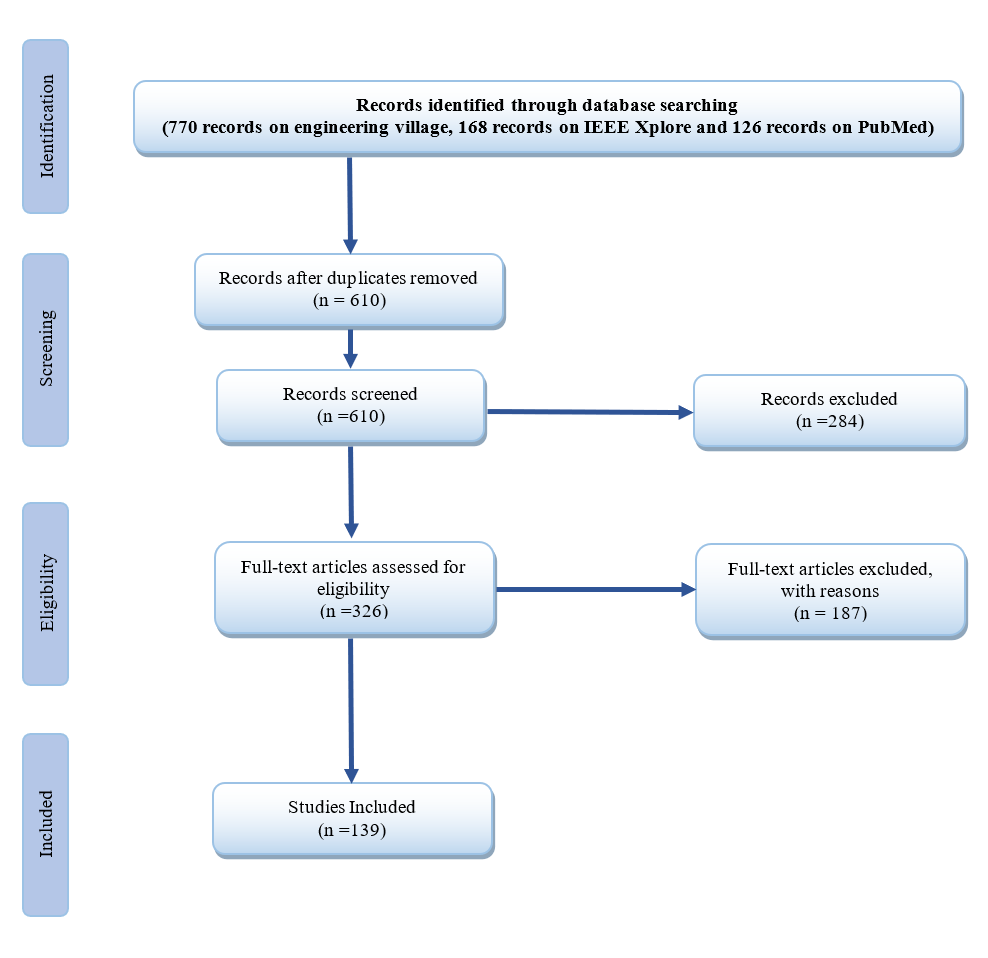}
\caption{PRISMA systematic review diagram \cite{Dressel2017}}
\label{fig:Prisma}
\end{figure*}

The screening procedure for the systematic review is based on evidence based PRISMA \cite{Moher2009} systematic and meta-analysis. The PRISMA diagram is shown in Figure \ref{fig:Prisma}. The research process was conducted in PubMed, IEEE Xplore, and Engineering Village, an engineering database with access to twelve additional databases including Ei Compendex, Inspec, GEOBASE, EnCompassLIT, USPTO Patents, Chemical Business NewsBase (CBNB),  EPO Patents, National Technical Information Service (NTIS), Geo-ref, PaperChem, EnCompassPAT, and Chimica \cite{Dressel2017}.
The search terms broadly involve two terminologies: 'Depression' and 'EEG' and 'machine learning'. Other inclusion criteria were conference articles, journal articles, open access and conventional (non-open access). A total of 123 publishers (on IEEE Xplore and Engineering Village) were included in our inclusion criteria to select papers. Some of the publishers are IEEE, Springer, Elsevier, World scientific, IOP publishing, Arxiv, Association for Computing Machinery, Scitepress, ASME, Nature Publishing Group, Frontiers, Hindawi, and Academic Press. In addition, PubMed has reported 30,000 records in their journal list \cite{NLM-AddedInfo}. 770 records were discovered in the Engineering Village database, 168 records were identified on IEEE Xplore and 126 records were also found in PubMed. 
After the duplicates were removed, 610 articles were detected for screening. Some of the papers were unavailable to download for the following reasons: (a) They were only abstracts and the main paper was missing from the journal, 
(b) The papers were focused on other mental health diseases and remotely connected to depression. Papers with the following subjects have been removed: schizophrenia only, ADHD, autism, stress, Sleep, LSTM, emotional valence, Hearing, Epileptic Seizures, Electroconvulsive, Reward, Response to Sertraline and Placebo Treatment, obsessive compulsive, antidepressant response, anesthetized, anxiety, consciousness, borderline personality disorder, Parkinson, gaming, treatment, rtms. After filtering, 139 articles were chosen for in-depth analysis and included in the systematic review. 

\subsection{Machine Learning Methodologies}
Below is a brief overview of the key machine learning methodologies. 

Supervised learning: This ML approach is used when the output is well defined. In this approach, the model is trained using labeled data to identify patterns and make predictions. Supervised learning can be divided into two categories, classification and regression. In classification, the model focuses on categorical predictions, while in regression, the model predicts continuous values. Popular supervised algorithms are logistic regression, linear regression, classification and regression trees (CART), Naive Bayes, neural networks, k-nearest neighbors (KNN) and support vector machines (SVM) \cite{chinnamgari2019r}. 

Unsupervised learning: This approach is used when the labeled data is unavailable. Unsupervised learning identifies patterns in the data to organize them. Unsupervised learning can be divided into clustering and association \cite{cios2007unsupervised}. k-means, k-modes, hierarchical clustering, and fuzzy clustering are popular clustering algorithms in unsupervised learning.

Semi-supervised learning: This method combines both supervised and unsupervised approaches and requires a large amount of data to train the ML model. In applications such as medical imaging, where a large amount of unlabeled data are available, the semi-supervised approach uses the small amount of labeled data to build a model that can label the large dataset \cite{cios2007unsupervised},\cite{chinnamgari2019r}. Popular Semi-supervised learning methods are Generative adversarial networks (GANs), semi-supervised support vector machines (S3VMs), graph-based methods, and Markov chain methods.

Reinforcement learning: this method is distinct from both supervised and unsupervised learning. Reinforcement learning is based on a reward-based system in which an agent optimizes its actions based on the feedback it receives. Through this process, the algorithm learns optimal solutions to maximize cumulative rewards over time. Popular reinforcement learning algorithms are Q-learning, SARSA, deep Q network (DQN), and deep deterministic policy gradient (DDPG).

\subsubsection{Machine Learning Algorithms}
The following is a brief introduction to  Machine learning algorithms:

Support Vector Machine (SVM): SVM is a supervised machine learning algorithm used for classification and regression tasks by constructing an optimal hyperplane that maximizes the margin between different classes \cite{Cortes1995}, \cite{Singh2016}. It is mainly effective in high-dimensional spaces and supports non-linear classification through kernel functions (e.g., Radial Basis Function (RBF), polynomial, sigmoid). Although SVM can handle high-dimensional feature spaces efficiently, performance may degrade if redundant or irrelevant features are included. Training SVM models, especially with non-linear kernels, can be computationally intensive due to quadratic programming complexity. The performance of SVM depends heavily on hyperparameter tuning, including the choice of kernel, regularization parameter CC, and kernel-specific parameters.

K-Nearest Neighbors (KNN): The KNN algorithm is a non-parametric, instance-based learning method used for classification and regression \cite{Zhang2016}. It classifies new samples by identifying the k nearest labeled data points and assigning the most frequent class among them (majority voting). Distance is typically measured using Euclidean distance. Since KNN does not construct an explicit model, it requires storing the entire dataset and performing real-time distance computations, making it computationally expensive for large datasets. KNN is sensitive to the presence of irrelevant or redundant features, which can distort distance calculations and degrade accuracy. Optimizations such as KD-Trees and Ball Trees can improve computational efficiency, particularly for high-dimensional data \cite{Zhang2016}.

Decision Trees: Decision trees are a widely used machine learning algorithm for classification and decision-making \cite{Podgorelec2002}. The algorithm systematically divides data into smaller subsets based on specific attribute values, forming a tree-like structure where internal nodes represent decision points and leaf nodes indicate final outcomes. One of the main strengths of decision trees is their simplicity, as they can handle both categorical and numerical data while offering a clear visual representation of decision processes. However, traditional algorithms can be prone to overfitting and may struggle with noisy or incomplete data.

Random Forests: Random Forest is an ensemble learning algorithm that enhances the predictive performance of individual decision trees by aggregating multiple tree models \cite{Au2018}. It operates using bootstrap aggregating (bagging), where multiple decision trees are trained on randomly sampled subsets of the data, reducing variance and mitigating overfitting. The method efficiently handles high-dimensional data, missing values, and both numerical and categorical predictors. However, categorical variables introduce challenges such as the "absent levels" problem, where unseen categorical values at inference time can lead to undefined splits, impacting model robustness. Techniques such as surrogate splits and heuristics for missing data are used to mitigate this issue.

Logistic regression: Logistic regression is a statistical and machine learning model for binary classification \cite{Hosmer2013},\cite{Levy2020}. It estimates the probability of an outcome given a set of predictor variables by applying the logit function, ensuring that outputs remain between 0 and 1. The model is optimized using maximum likelihood estimation (MLE), and performance is typically assessed using goodness-of-fit tests, AUC-ROC, and classification metrics. While logistic regression assumes a linear relationship between predictors and log-odds, it can be extended with interaction terms or regularization techniques like LASSO and Ridge regression to improve generalization. Although machine learning models like random forests often outperform logistic regression in high-dimensional and complex datasets, logistic regression remains valuable for its interpretability and statistical inference capabilities.

Neural networks (NN): Neural Networks are computational models designed to recognize patterns and make predictions by mimicking the structure of biological neurons. They consist of interconnected layers: an input layer that receives raw data, hidden layers that extract features, and an output layer that produces the final result. Learning occurs through weight adjustments using optimization algorithms such as gradient descent, allowing NNs to generalize data \cite{Goodfellow-et-al-2016}. NNs have advantages such as adaptability to complex patterns, efficient feature learning, and scalability for large datasets. However, they require large datasets, significant computational power and are prone to overfitting without proper regularization. 

Various types of NNs have different applications. Feedforward Neural Networks (FNNs) process data in one direction and are commonly used in classification and regression. Recurrent Neural Networks (RNNs) introduce feedback loops, making them suitable for sequential tasks like time-series prediction and speech recognition. However, traditional RNNs struggle with long-term dependencies, a challenge addressed by Long Short-Term Memory (LSTM) networks, which maintain memory over extended sequences \cite{mandic2001}.Convolutional Neural Networks (CNNs) are specialized for image and video processing, using convolutional layers to detect spatial patterns.

Autoencoders, composed of encoder-decoder structures, are useful for dimensionality reduction and anomaly detection. Generative Adversarial Networks (GANs) generate synthetic data by training two competing networks, finding applications in image synthesis and style transfer \cite{Goodfellow-et-al-2016}. Applications of NNs span multiple domains. FNNs are used in medical diagnosis and fraud detection, RNNs and LSTMs in speech recognition and financial forecasting, and CNNs in object detection and medical imaging. Autoencoders support data compression, while GANs enhance artificial content generation.\cite{mandic2001}, \cite{Goodfellow-et-al-2016}.

\section{Review of algorithms to diagnose depression}
\subsection{EEG data acquisition}

All studies in our systematic review have used an EEG dataset for their analysis. 102 of 139 (74. 38\%) studies have collected their own EEG data, while only 20 research groups used a dataset that belonged to another study. The participant selection criteria varied between different research groups; 106 out of 139 (76. 25\%) studies used a screening procedure to choose participants. The selection criteria for the choice of subjects were clinical interviews, questionnaires, and rating scales. 28 of 139 research studies did not mention the selection criteria they used to choose patients. 
120 of 139 (86.33\%) experiments were carried out in a controlled environment. Depending on the experiment in each research study, the EEG data was recorded while subjects were in the resting state with their eyes open or closed and with audio and visual stimulation. This is given in Tables \ref{tab:Table3-11} to \ref{tab:Table3-5}. 12 of 139 research groups did not mention the condition of the experiment. The duration of the analysis was also significantly different in each study and ranged from 20 seconds to 34 minutes and in one experiment, from 9 to 10 hours. This was indicated by 118 of the 139 (84.89\%) studies. 21 of 139 research studies did not provide any information on the analysis period.
In this systematic review, the summary of the most recent literature based on the number and type of electrodes, their position, data acquisition device, sampling frequency, impedance and type of filters have also been provided in Tables \ref{tab:Table4-1} to \ref{tab:Table4-155}. 96 of 139 (69. 06\%) papers provided details of their data collection systems, while 43 of 139 papers did not mention any information about their data collection systems. 127 of 139 (91. 36\%) papers mentioned the number of electrodes they used for their analysis. The minimum number of electrodes to collect data was 2 and the maximum number of electrodes was 128. In some of the papers, single electrode EEG diagnosis was studied, although the data was collected with a larger number of electrodes.

\subsection{Preprocessing and feature extraction }

Similar to many applications of EEG-based biomarkers, in most of the studies in our systematic review, preprocessing pipelines were implemented to remove artifacts from the EEG signal and denoise the signal. Of 139 papers, 43 studies (30.93\%) have only used a bandpass filter, while 26 papers (18.7\%) have used a combination of a bandpass and notch filter. 10 of 139 papers (7.19\%) have also utilized a combination of low-pass, high-pass, and notch filter and 6 studies (4.31\%) have only used low-pass and high-pass filter.  
 Eight papers (5.7\%) of studies have reported that only a notch filter has been used to remove powerline noise and its harmonies, and only one paper has used a combination of a high-pass and notch filter (0. 71\%). 3 out of 139 studies (2.15\%) have reported that a low pass filter has been used and another 2 studies have also mentioned the use of only a high pass filter. The authors of 4 out of 139 studies (2. 87\%) have used a combination of wavelet filters, and another study has used a rectangular moving average filter. In addition, 2 studies have used a digital filter with cut-off frequencies of fc = 0.5 Hz and fc = 40 Hz, and another study has used median and band-stop filters. 
27 out of 139 (19.42\%) of the studies have not mentioned any traditional filtering stage namely \cite{Li2022},\cite{Garg2022}, \cite{Sharmila2022}, \cite{Lin2022}, \cite{Thakare2022}, \cite{Loh2021}, \cite{Hong2021}, \cite{Duan2021}, \cite{Zhu2020}, \cite{Saleque2020}, \cite{8101151}, \cite{Liao2013}, \cite{8981929}, \cite{Ke2019}, \cite{Mumtaz2019}, \cite{19258582} ,\cite{18853691}, \cite{20184806131498} , \cite{20173504090707}, \cite{Wan2017AQA}, \cite{Wan2017},\cite{17051427}, \cite{Frantzidis2012}, \cite{20160801982396} , \cite{20070710426068}, \cite{Pockberger1985} and \cite{20185106271469}. One of the reasons that the authors have not reported the filtering characteristics is that in their experiments the filtering stage is integrated in their data acquisition system hardware.
In addition to basic spectral filtering, more advanced artifact removals have also been utilized. Independent component analysis (ICA) was extensively used by several studies \cite{Movahed2022}, \cite{Zhao2022}, \cite{Kim2022}, \cite{Sun2022}, \cite{Babu2022}, \cite{Seal2022}, \cite{Ghiasi2021}, \cite{Movahed2021}, \cite{Li2021}, \cite{Sharma2021}, \cite{Uyulan2021}, \cite{Wang2021}, \cite{Seal2021}, \cite{Shen2021}, \cite{Duan2020}, \cite{Apsari2020}, \cite{Saleque2020}, \cite{Thoduparambil2020}, \cite{Mahato2020}, \cite{Mahato2020_2}, \cite{Kang2020}, \cite{20193307311994}, \cite{18951307}, \cite{19258582}, \cite{18927189}, \cite{20183405723339}, \cite{20185106271469}, \cite{20183605764415}, \cite{XiaoweiLi2016EEG-basedClassifiers}, \cite{Spyrou2016}, \cite{Frantzidis2012}, \cite{Ku2012}. Wavelet-based techniques have also been employed by several papers as a preprocessing algorithm to remove artifacts \cite{Song2022}, \cite{WeiEEGBASED2021}, \cite{Saeedi2020}, \cite{19203658}, \cite{18853691}, \cite{Cai2018ACASE}, \cite{Cai2018ADetection}, \cite{Wan2017AQA}, \cite{Shen2017ACriterion}, \cite{Zhao2017}, \cite{Wan2017}, \cite{Cai2017NO2}, \cite{Puthankattil2017}, \cite{Cai2016PervasiveCollector}. Most of the papers extracted linear and non-linear features from their dataset and created a feature space for further analysis. A short description of the type of features has been mentioned, and further details of the features are shown in Tables \ref{tab:Table5-1} to \ref{tab:Table5-4}.

\subsection{Data processing and machine learning}
In this section, we report a summary of each article based on machine learning algorithms and statistical analysis utilized. Some of the authors have used a combination of different data processing techniques and concluded that one of them had achieved the highest performance metrics. This is mentioned in Tables \ref{tab:Table2-1} to \ref{tab:Table2-17}. In addition, performance metrics (accuracy, specificity and sensitivity) of the techniques have also been reported in Tables \ref{tab:Table2-1} to \ref{tab:Table2-17}. Some of the authors have not reported performance metrics or only reported the accuracy of the results in their articles. This is reported as N/A in the tables. In the next section, we have categorized the papers with highest accuracy based on the data processing technique and the machine learning approach that they have used.

\subsubsection{Logistic Regression}
16 out of 139 papers (11.51\%) of papers we studied used logistic regression in their analysis. Hosseinifard et al. \cite{Hosseinifard2013} explored a nonlinear method to characterize depressed and healthy subjects. In their study, they used EEG signals from 19 channels located on prefrontal, frontal, parietal, central, temporal and occipital lobes to extract non-linear and linear features, including Lyapunov exponent (LE), correlation dimension, Higuchi fractal (HF), and band powers. The authors also employed detrended fluctuation analysis (DFA) to calculate correlation properties and fractal scaling of the EEG signal. Features were selected using GA. LR, LDA, and KNN classifiers were used to classify features. Among linear features, Alpha band power and among nonlinear features, correlation dimension showed highest accuracy to discriminate patients. The results show that Logistic Regression (LR) classifier can achieve higher accuracy in comparison to KNN and LDA.
In a study by W. Mumtaz et al. \cite{Mumtaz2015ADisorder}, the authors used the choice of three EEG references using LR and SVM to discriminate patients with depression from healthy controls. 33 depressed and 19 healthy subjects participated in their study. The EEG signal was collected from 19 electrodes using BrainMaster 24 E amplifier (BrainMaster Technologies, Inc., USA) at the sampling frequency of 256 Hz. The EEG data was denoised with Surrogate filtering technique. Inter-hemispheric asymmetries, Power of different EEG frequency bands and coherenc were exracted from the signal. The authors studied three EEG references of average reference (AR), infinity reference (IR), and link-ear (LE) reference and achieved the highest accuracy, specificity, and sensitivity with LR. In addition to above papers, authors in \cite{Babu2022}, \cite{Jan2022}, \cite{Li2021}, \cite{Jiang2021}, \cite{Hong2021}, \cite{WeiEEGBASED2021}, \cite{Duan2021}, \cite{Saleque2020}, \cite{Sun2020}, \cite{Mahato2020_2}, \cite{Ding2019}, \cite{Cai2018StudyFS} and \cite{Li2015Second} have also used Logistic Regression in their studies.

\subsubsection{Support vector machine (SVM)}
In our systematic review, 53 out of 139 papers (38.12\%) used the SVM classification algorithm to classify depressed patients from healthy control subjects. I. Kalatzis et al. \cite{8101151} were one of the first research groups to use SVM with the Majority-Vote engine to identify patients with depression. In their study, 25 depressed and 25 healthy subjects were selected. The authors used audio stimulation to collect resting-state EEG data from 15 Ag/AgCl electrodes at the sampling frequency of 500 Hz. 17 features were extracted from EEG signal and the authors achieved the maximum classification accuracy with all the leads. During specific tasks, right brain hemisphere dysfunction was observed in depressed group.
T.Tekin Erguzel et al. \cite{TekinErguzel2015} proposed an algorithm based on Ant Colony Optimization (ACO) and SVM. EEG data were collected from 55 patients with mild depression and 46 patients with bipolar disorder. Neuroscan EEG cap (Compumedics, NC, USA) with 19 Ag/AgCl electrodes were used to collect EEG data at the sampling frequency of 250Hz. The authors proposed an Improved ACO technique to select 22 of 48 features. In their research, they achieved the highest performance metrics with the IACO-SVM algorithm. The authors calculated the Area under curve (AUC) of 0.793. Using nested cross-validation (CV), it was shown that the accuracy of the proposed algorithm was higher than Particle swarm optimization (PSO), Genetic algorithm (GA), and ACO algorithm. 
K. M. Puk et al. \cite{Puk2016} explored a technique to discriminate depressed and healthy subjects based on the effects of depression on memory processing. 15 depressed and 12 normal subjects participated in the study, and EEG data were collected with a 32 channel Brain vision system from (Brain products GmbH, Germany) at a sampling frequency of 1000Hz. In order to reduce the size of the dataset, the signal was resampled to 256Hz, and 9 groups of linear and nonlinear features were extracted from the signal. In order to select features with the highest discriminative properties, the Minimal-Redundancy Maximal-Relevance criterion (mRMR) technique was employed. The selected features were classified with the SVM classifier, and the classification accuracies of over 80\% were achieved. 
In another study by J. Shen et al. \cite{Shen2017ACriterion}, a method was developed by the authors to diagnose depression with three Fp1, Fp2, and FpZ electrodes. EEG data were collected from 81 Depressed and 89 healthy subjects with a Three-electrode pervasive EEG collection device (Ubiquitous Awareness and Intelligent Solutions (UAIS), Lanzhou University, China) at the sampling frequency of 250Hz. Ocular artifacts were removed by Discrete wavelet transform (DWT) and Kalman filter. Three linear features of Centroid frequency, Mean frequency and Max frequency and three nonlinear features of C0 complexity, Correlation dimension and Renyi entropy were extracted from five EEG frequency bands: Alpha (8-13 Hz), Beta (13-30Hz), Theta (4-8 Hz), Delta (1-4 Hz) and Full band (1-40Hz). The authors used SVM to discriminate depressed and healthy subjects.
W. Mumtaz et al. \cite{20172903942647} proposed a machine learning algorithm based on Synchronization likelihood (SL) to discriminate depressed and healthy subjects. In this study, EEG data were collected from 34 depressed and 30 control subjects using 19 electrode EEG cap with a BrainMaster Systems amplifier (BrainMaster technologies Inc., USA) at sampling frequency of 256Hz. The authors used the Multiple Source technique to remove artifacts. A feature matrix was created for each subject, and a rank-based algorithm was used to select relevant features. Selected features were classified with 10-fold cross-validation with Logistic Regression (LR), SVM, and Naive Bayesian (NB) classifiers. The highest performance metrics was achieved with SVM classifier. The authors claimed that the 95\% specificity of the proposed method shows that the algorithm can potentially be used in clinical setting. 
I. Spyrou et al. \cite{Spyrou2016} studied neurophysiological features of 34 patients with Geriatric depression and neurodegeneration with 32 healthy subjects to discriminate between the two groups. A Nihon Kohden JE-207A (Nihon Kohden Europe GmbH) data acquisition system was used to collect EEG data. The subjects asked to wear a 57 electrode cap (EASYCAP from EasyCap GmbH) and data was collected at sampling frequency of 500Hz. In their approach, they extracted oscillatory and synchronization features based on Discrete wavelet transform (DWT). The authors employed the Random Forest (RF), Random Tree (RT), SVM and Multilayer Perceptron (MLP) classifiers to calculate the accuracy of their method. They achieved the highest classification accuracy, specificity and sensitivity with the RF classifier. 
S. Mantri et al. \cite{15670461} used EEG linear analysis and SVM classifiers to discriminate 13 depressed and 12 control subjects. EEG signals were captured with an 8-channel data acquisition device with F3, F4, Fz, C3, C4, Pz, P3, and P4 electrodes which are located at Frontal, Occipital and Parietal lobes. The sampling frequency of 256 Hz was used for data acquisition. The power spectrum of four different bands was calculated, and the extracted features were classified with SVM classifier. The results showed that the proposed SVM classifier can be utilized to achieve the highest performance metrics.
X. Li et al \cite{20193307311994} studied the reliability of EEG machine learning analysis based on emotional face stimulation task to discriminate depressed and healthy subjects. 14 depressed and 14 healthy subjects participated in the experiment. The authors used a 128 electrode HydroCel Geodesic Sensor Net (Electrical Geodesics, Inc., USA) with 250 Hz sampling frequency to collect EEG data. 16 electrodes located on prefrontal, frontal, central, temporal, parietal and occipital lobes were selected for analysis. FastICA algorithm and an adaptive noise canceller based on LMS algorithm were employed to remove artifacts. Power spectral density and activity and Hjorth features were extracted from the EEG data and the authors employed an ensemble method based on deep forest and SVM. The best performance metrics were achieved with ensemble model and power spectral density. 
In another study by S.C. Liao et al \cite{Liao2017}, an a new feature extraction method based on spectral-spatial EEG features was proposed to discriminate depressed and healthy subjects based on. 12 depressed and 12 healthy subjects were participated in the experiment. EEG data was collected from 30 Ag/AgCl electrodes with a Quick-Cap 32 EEG data acquisition device at 500 Hz sampling frequency. The ocular artifacts were removed by NeuroScan artifact removal software. Theta (4–8 Hz), alpha (8–13 Hz), beta (13–30 Hz) and gamma (30–44 Hz) band powers and correlation dimension were extracted from the EEG signal. The highest performance achieved with the combination of kernel eigen-filter-bank common spatial pattern (KEFB-CSP) feature extraction method proposed by authors and SVM classifier. The authors achieved the highest accuracy, specificity and sensitivity of with 8 electrodes and 6 seconds of recorded EEG data.
Authors in \cite{Mumtaz2017} proposed a machine learning platform to discriminate depressed patients and healthy subjects. The researchers collected EEG data from 33 depressed and 33 healthy subjects. To collect data, they used 19-channels EEG cap with BrainMaster Systems amplifier (BrainMaster technologies Inc., USA) at the sampling frequency of 256 Hz. The electrodes were located on frontal, temporal, parietal and occipital lobes. Multiple source techniques with standard brain electric source analysis (BESA) software was used to remove artifacts and alpha interhemispheric asymmetry and EEG spectral power features were extracted from EEG data. The proposed machine learning platform achieved the highest performance metrics with SVM classifier. 
In another study by M. Sharma et al. \cite{18983501}, the authors proposed a method to diagnose depression based on  three-channel orthogonal wavelet filter bank (TCOWFB). 15 depressed and 15 healthy participated in the study. The EEG data was collected from Bipolar channels FP2-T4 (right half) and Fp1-T3 (left half) at 256 Hz sampling frequency. Ocular and muscle artifacts were manually removed with visual inspection and nonlinear features were extracted from seven wavelet sub-bands. The authors concluded their method has lower computational complexity and higher performance in comparison to previous studies.
In addition to the mentioned papers, authors of other papers including \cite{Li2022}, \cite{Zhao2022}, \cite{Kim2022}, \cite{Zhang2022}, \cite{Lin2022}, \cite{Avots2022}, \cite{Sun2022}, \cite{Nayad2022}, \cite{Babu2022}, \cite{Jan2022}, \cite{Seal2022}, \cite{Ghiasi2021}, \cite{Movahed2021}, \cite{Movahed2021}, \cite{Li2021}, \cite{Sharma2021}, \cite{Jiang2021}, \cite{Hong2021}, \cite{WeiEEGBASED2021}, \cite{9477800}, \cite{Shen2021}, \cite{Duan2021}, \cite{Saeedi2020}, \cite{Duan2020}, \cite{Saleque2020}, \cite{Mahato2020}, \cite{Mahato2020_2}, \cite{Ding2019}, \cite{Cai2018StudyFS}, \cite{Li2015Second}, \cite{Orgo2017}, \cite{17991812}, \cite{20193407357198}, also used SVM algorithm in their studies. Details of the utilized techniques are provided in Tables \ref{tab:Table2-1} to \ref{tab:Table2-17}. 
\subsubsection{K-Nearest Neighbor (KNN)}
In addition to the SVM classifier, 25 out of 139 papers (17.98\%) of papers have used the KNN classifier in their studies. 
X. Li et al. \cite{15798207} proposed an algorithm using linear and nonlinear features with the KNN classifier. EEG data was collected from 9 depressed and 25 control subjects with a 128 channel Geodesic sensor net (Electrical Geodesics, Inc., USA). The sampling frequency was set to 250HZ. 20 linear and nonlinear features including sum power, max power, variance, Co-complexity, Lyapunov, and Kolmogorov entropy were extracted from delta (0.5–4Hz), theta (4–8Hz), alpha (8–13Hz) and beta (13–30Hz) frequency bands. The researchers implemented methods based on KNN, SVM, LR, NB and RF classifiers. The authors achieved the classification accuracy of over 99\% based on KNN classifier with the combination of linear and nonlinear features.
 In another study by X. Zhang et al. \cite{13386381}, the authors proposed an approach to characterize depressed and healthy females with three electrodes (Fp1, Fp2, and FpZ) on prefrontal lobe. The researchers developed a 3-electrode mobile EEG belt for EEG data acquisition and collected data at the sampling frequency of 256 Hz. In their method, several linear and nonlinear features were extracted from 13 depressed and 12 healthy subjects. KNN, and Backpropagation neural network (BPNN) classifiers were used to classify depressed and control subjects. In their study, the authors achieved higher classification accuracy with BPNN in comparison to KNN.
X. Li et al. \cite{XiaoweiLi2016EEG-basedClassifiers} proposed an algorithm based on Greedy Stepwise (GSW) and KNNs to discriminate patients with mild depression and healthy subjects. The authors used a 128 channel HydroCel Geodesic Sensor Net (Electrical Geodesics, Inc., USA) at the Sampling frequency of 250 Hz to collect data from 10 depressed and 10 control subjects. Eight linear and 9 non-linear features were extracted, and five feature selection methods of Greedy Stepwise, Best First (BF), Linear Forward Selection (LFS), Genetic Search (GS), and RankSearch (RS) were used for data dimensionality reduction. The authors achieved the highest classification accuracy with the GSW-KNN algorithm using 16 electrodes. It was concluded that five electrodes of T3, O2, Fp1, Fp2, and F3 which are located on temporal, occipital, prefrontal and frontal lobes could achieve high performance metrics.  
S. Zhao et al. \cite{Zhao2017} designed a real-time system to monitor and discriminate depressed patients and control subjects. The EEG data were collected with three Fp1, Fp2, and FpZ electrodes which are located on prefrontal lobe at the sampling rate of 256Hz with an audio stimulus. The authors used a Finite impulse response (FIR) filter, a mid-filter, Wavelet transforms, and Kalman filter to remove noise. Three linear features of Max frequency, Mean Frequency, and Center Frequency in combination with three non-linear features of C0 Complexity, Permutation Entropy and Lempel Ziv complexity (LZC) were extracted from the signal. The proposed algorithm using six extracted features achieved and average classification accuracy of over 78\% with Local classification based on KNN and Naïve Bayes. 
H. Cai et al. \cite{Cai2018ADetection} explored the feasibility of developing an algorithm to differentiate depressed patients with three electrodes. The authors collected data from 92 depressed and 121 control subjects with data acquisition device (Ubiquitous Awareness and Intelligent Solutions (UAIS), Lanzhou University, China) using three electrodes of Fp1, Fp2, and FpZ at the sampling frequency of 250 Samples per second. To denoise the signal, the authors used DWT, Adaptive predictor filter (APF) and FIR filters. Two hundred seventy linear and nonlinear features were extracted from the signal, and the authors used the Minimal-Redundancy-Maximal-Relevance algorithm to select relevant features from the feature matrix. The highest classification accuracy was achieved with KNN. It was concluded that the absolute power of theta could be a significant indicator to discriminate depressed and healthy controls. 
In another study by H. Cai et al. \cite{Cai2018ACASE}, the authors proposed a new method based on normalized Euclidean distance to increase the accuracy of the previous algorithm. 86 depressed and 92 healthy subjects were participated in the study and data was collected from three Fp1, Fp2, and FpZ electrodes on prefrontal lobe with a data acquisition device (Ubiquitous Awareness and Intelligent Solutions (UAIS), Lanzhou University, China) at the sampling frequency of 250 Hz. Wavelet transform was used to remove signal artifacts and a combination of linear and nonlinear features were extracted. The authors achieved the highest accuracy with the KNN classifier with the proposed method.
In addition to the selected papers summarized in the above, authors in \cite{Kim2022}, \cite{Zhang2022}, \cite{Avots2022}, \cite{Nayad2022}, \cite{Babu2022}, \cite{Seal2022}, \cite{Li2021}, \cite{Jiang2021}, \cite{WeiEEGBASED2021}, \cite{9477800}, \cite{Saeedi2020}, \cite{Duan2020}, \cite{Sun2020}, \cite{Cai2018StudyFS}, \cite{Li2015Second} and \cite{19203658} have also used KNN classifier as mentioned Tables \ref{tab:Table2-1} to \ref{tab:Table2-17}..

\subsubsection{Neural Networks}

44 out of 138 papers (31.65\%) of papers we studied used neural networks in their analysis. S. D. Puthankattil et al. \cite{Puthankattil2012ClassificationEntropy} employed Relative Wavelet Energy (RWE) and Artificial Neural Network (ANN) classifiers to discriminate 30 healthy and 30 depressed subjects. They used Fp1-T3 and Fp2-T4 electrodes to collect EEG data at the sampling frequency of 256 Hz. In their study, Ocular and muscle artifacts were manually removed, and Total Variation Filtering (TVF) were employed to denoise the signal. They found the signal energy distribution can be used to characterize depressed and control subjects.
Y. Katyal et al. \cite{Katyal2014} proposed an algorithm based on the combination of EEG signal and facial emotion analysis. In their study, 10 healthy and 10 depressed subjects were participated. The data was collected with 64 EEG electrodes at the sampling frequency of 256 Hz. They used ICA to eliminate artifacts and Wavelet packet decomposition (WPD) was employed to extract features from EEG data. To classify features from EEG signals and facial recognition analysis, the authors proposed a classifier based on the ANN. The authors concluded they could achieve higher accuracy by combining EEG analysis and facial recognition.
T. Erguzel et al. \cite{Erguzel2016} proposed an approach based on cordance values (which is calculated based on quantitative electroencephalographic (QEEG)), PSO and ANN. The EEG data were collected from 31 bipolar and 58 unipolar subjects with 19 Ag/AgCl electrodes using Scan LT EEG Amplifier (Compumedics/Neuroscan, USA) with an EEG cap at the sampling frequency of 250 Hz. The authors then used normalized absolute and relative powers of each electrode to calculate cordance values for each frequency band. PSO evolutionary computation algorithm was used to select features from the feature matrix. The proposed approach based on hybrid PSO and ANN achieved the highest performance metrics. 
Y. Mohan et al. \cite{Mohan2016} presented an algorithm based on ANN to classify two groups of 5 depressed and 5 control subjects. Data was collected with a 32 wet electrode data acquisition device at the sampling frequency of 128Hz. A multilayer feed-forward network was used to classify data using 800 input neurons per electrodes. The authors concluded that C3 and C4 electrodes in the central part of the brain can be utilized to classify depressed and healthy subjects with high accuracy.
H. Cai et al. \cite{Cai2016PervasiveCollector} studied the feasibility of a portable EEG system to discriminate depressed and healthy subjects based on Deep Belief Networks (DBN). 86 depressed and 92 control subjects participated in the experiment. The researchers used a Three-Electrode Pervasive EEG Collector (Ubiquitous Awareness and Intelligent Solutions (UAIS), Lanzhou University, China). The electrodes were Fp1, Fp2, and FpZ, which are located on the prefrontal cortex. 28 linear and nonlinear features were extracted and KNN, SVM, ANN and DBN were employed to analyze features and find features with the highest discriminative power. The results indicate that DBN and the absolute power of Beta wave (13-30Hz) can achieve the highest accuracy.
S. D. Puthankattil et al. \cite{Puthankattil2017} studied the Probabilistic Neural Network (PNN) and Feedforward Neural Network (FFNN) methods to discriminate depressed and healthy subjects. EEG data were collected from 30 depressed and 30 control subjects with a 24-channel data acquisition system at 256Hz sampling frequency. The authors used DWT to decompose the signal and extracted time-domain features. In addition, RWE and wavelet entropy (WE) were also calculated. The authors achieved highest classification accuracies with RWE-FFNN and WE-FFNN respectively. It was concluded the proposed algorithm based on time and domain features with FFNN could achieve higher classification accuracy compared to other classifiers. 
U. Acharya et al. \cite{Acharya2018AutomatedNetwork} proposed a technique to classify depressed and healthy subjects based on Convolutional Neural Network (CNN). The proposed algorithm can discriminate patients and healthy controls without using a feature matrix. The authors used Fp1-T3 and Fp2-T4 channel pairs to collect data from 15 depressed and 15 healthy subjects at the sampling frequency of 256Hz. It was concluded that the right hemisphere EEG signals could achieve higher classification accuracies compared to left hemisphere EEG signals. 
Y. Guo et al. \cite{20173504090707} proposed a classification function based on linear discriminant analysis (LDA) and a new algorithm based on the Multi-Objective Particle Swarm Optimization (MOPSO) algorithm to discriminate depressed and healthy subjects. The authors used a 14-electrode data acquisition system with Emotive headset (Emotive Inc., USA) at 128 SPS (samples per second) to collect EEG data from 3 depressed and 3 healthy subjects. The authors used AF3, AF4, F3, F4, F7, F8, FC5, FC6, T7, T8, P7, P8, O1 and O2 electrodes to collect data. LDA and MOPSO classification results achieved the highest performance metrics with 6 volunteers.
H. Cai et al. \cite{Cai2017NO2} studied depression diagnosis based on gender differences in a group of 90 depressed and 68 healthy subjects. A Three-electrode EEG collector (Ubiquitous Awareness and Intelligent Solutions (UAIS), Lanzhou University, China) was used to collect EEG data. Two electrodes of Fp1 and Fp2 which are located on prefrontal lobe were selected for further analysis and ocular, muscle and powerline artifacts were removed by a bandpass filter and wavelet transform. The authors extracted 32 linear and nonlinear features from the signal and employed the Sequential Floating Forward Selection (SFFS) algorithm and SVM, KNN, DT and ANN classifiers to find the most effective features to discriminate depressed and healthy subjects. The highest classification accuracy to discriminate male subjects achieved with ANN classifier while for female subjects the highest classification accuracy observed with KNN. It was concluded that for male subjects the best results achieved under resting state and females, the highest classification accuracy was obtained under negative audio stimulation.
P. Sandheep et al. \cite{Sandheep2019} proposed a machine learning approach based on CNN to distinguish depressed and healthy subjects. 30 depressed and 30 healthy subjects were selected for the experiment and EEG data was collected from Left half (Fp1-T3) and right half (Fp2-T4) of the brain at 256Hz sampling frequency. After z-score normalization to normalize data, the authors employed a five-layer CNN. The results showed that the classification accuracy was higher in the right-brain hemisphere in comparison to left-brain hemisphere. The researchers concluded that with their proposed five-layer CNN platform and without using a feature extraction algorithm, high performance metrics to discriminate depressed and healthy subjects is achievable. 
In another study by W. Mumtaz et al. \cite{Mumtaz2019}, two deep learning depression diagnosis frameworks based on One dimensional CNN (1DCNN) and 1DCNN with long short-term memory (LSTM) were proposed. 33 Depressed and 30 Healthy subjects participated in the study and EEG data was collected from 19 channels at the sampling frequency of 256 Hz. To eliminate ocular artifacts, multiple source eye correction (MSEC) algorithm was employed. The highest performance metrics were achieved with 1DCNN model and 1DCNN with LSTM model respectively. It was concluded that the proposed 1DCNN models can be used in developing wearable systems due to low complexity and high performance of the models.
S. Mahato et al. \cite{20183405723339} explored linear, nonlinear and combination of features to distinguish depressed and healthy subjects. The authors selected 34 depressed 30 healthy for the experiment and using 19 electrodes they collected EEG data from frontal, temporal, parietal, occipital and central lobes of subjects at the sampling frequency of 256 Hz. ICA and common average reference (CAR) were employed to remove signal artifacts. Band power (delta (0.5–4 Hz), theta (4–8 Hz), alpha (8–13 Hz) and beta (13–30 Hz)) and Interhemispheric asymmetry linear features were extracted. In addition, nonlinear features of Wavelet transform, Relative wavelet energy (RWE) and Wavelet entropy (WE) were also extracted from the EEG signal. The researchers employed Principal component analysis (PCA) algorithm as a dimension reduction algorithm to reduce the number of irrelevant features. The EEG data was analyzed using Multi-layered perceptron neural network (MLPNN), radial basis function network (RBFN), LDA and quadratic discriminant analysis (QDA). The highest classification accuracy, specificity, and sensitivity was achieved with the combination of alpha power and RWE features with MLPNN and RBFN classifiers. 
In another study by H. Ke et al. \cite{20184806131498} , The authors proposed a cloud-based machine learning framework for wearable systems using a light weight CNN. The authors collected EEG data from three groups of healthy and pressed subjects. Dataset 1 consists of 34 depressed and 30 healthy subjects and dataset 2 consists of 17 depressed patients. The data was collected with 20 electrodes located on prefrontal, frontal, temporal, central, parietal and occipital lobes at the sampling frequency of 256 Hz. The proposed light weight CNN achieved classification accuracy of over 98. It was concluded that the proposed machine learning framework is suitable for use in wearable healthcare application due to high performance, fast processing time and low computational complexity.
O. Faust et al. \cite{Faust2014} proposed a PNN machine learning framework for depression treatment and diagnosis. The authors chose 30 depressed and 30 healthy subjects for their experiment and collected EEG data from Fp1-T3 (left half of brain) and FP2-T4 (right half of brain) at the sampling frequency of 256 Hz. Artifacts were removed by TVF algorithm and visual inspection. A combination of linear features of Wavelet packet decomposition and nonlinear features of Bispectral phase entropy, Renyi entropy, Sample entropy and Approximate entropy were extracted from the EEG signal. Student’s t-test was employed to select discriminative features and different classifiers such as KNN, SVM, DT, NB, Gaussian mixture model (GMM), Fuzzy Sugeno Classifier (FSC) and PNN were utilized to calculate the performance metrics. The authors achieved the highest performance metrics with PNN.
In addition to the selected papers summarized above, other articles \cite{Wang2DCNN2022}, \cite{Sharmila2022}, \cite{Zhang2022}, \cite{Song2022}, \cite{Wang2022}, \cite{Nayad2022}, \cite{Goswami2022}, \cite{Jan2022}, \cite{Seal2022}, \cite{Thakare2022}, \cite{Loh2021}, \cite{Savinov2021}, \cite{Uyulan2021}, \cite{Seal2021}, \cite{Duan2020}, \cite{Thoduparambil2020}, \cite{Kang2020}, \cite{8981929}, \cite{18768750}, \cite{19259295}, \cite{Ke2019}, \cite{18951307}, \cite{19258582}, \cite{18927189}, \cite{20185106271469}, \cite{20194907787800} also used Neural networks in their research, as shown in Tables \ref{tab:Table2-1} to \ref{tab:Table2-17}.

\subsubsection{Statistical Analysis}

31 out of 139 papers (22.30\%) in our systematic review have used statistical analysis to discriminate depressed patients and healthy subjects. 
The authors in \cite{20102413002260} studied Spectral Asymmetry (SA) with Mann-Whitney U-test and Bonferroni Correction on 18 Depressed and 18 Healthy female subjects. They collected EEG data with Cadwell Easy II (Cadwell Industries Inc., USA) EEG data acquisition system at the sampling frequency of 400 Hz. 19 electrodes located on frontal, parietal, temporal and occipital lobes were used during the experiment. Power spectral density (PSD), EEG band Powers and SA features were extracted from the EEG signal. Their results show that the proposed technique can distinguish depressed and control subjects based on SA and confirm higher relative beta and beta power in depressed patients in all the areas of the scalp. 
Authors in \cite{12424821} explored cortical functional connectivity of 12 depressed and 12 healthy subjects by using a method of partial directed coherence (PDC). In their research, they used a 16-channel data acquisition system (Sunray, LQWY-N, Guangzhou, China) with 100Hz sampling frequency. They placed the electrodes on prefrontal, frontal, central, parietal, occipital and temporal regions of brain. The authors removed the artificats manually. One-way ANOVA statistical analysis showed the presence of Hemispheric Asymmetric Syndrome in depressed patients.
M. Bachmann et al. \cite{Bachmann2015} studied the complexity of EEG signals using Lempel Ziv Complexity method. They used an 18 channel Cadwell Easy II (Cadwell Industries Inc., USA) EEG data acquisition system with 400Hz sampling frequency to collect EEG data from 17 depressed and 17 control subjects. Electrodes were located on prefrontal, frontal, central, parietal, temporal and occipital lobes. Mann-Whitney statistical test showed a significant difference in the complexity of the algorithm between two groups.
Z.Liao et al. \cite{Liao2013} studied correlation and Asymmetry analysis on 10 healthy, 7 unmedicated, and 5 medicated depressed patients taking antidepressant. They used six forehead and parietal electrodes (Fp1, Fp2, F3, F4, P3, and P4) with the Brain Vision Recorder (Brain products GmbH, Germany) at the sampling frequency of 500 Hz. To remove artifacts, the authors used Brain Vision Recorder (Brain products GmbH, Germany) and Analyzer Filter. The proposed method proved the correlation between the beta frequency band in the frontal brain area and the (Beck Depression Inventory) BDI score.
M. Sun et al. \cite{Sun2015} studied the power spectral analysis to categorize depressed and healthy patients. The authors collected EEG data from 11 healthy and 11 depressed patients with three prefrontal electrodes (FP1, Fp2 and FpZ) at the sampling frequency of 256Hz. Gravity frequency of power spectrum and relative power features were extracted from the signal. The Results from statistical analysis showed depressed patients have lower relative power in the alpha band (8-14Hz) and higher relative power in the beta band (14-30Hz).
D. P. X Kan et al. \cite{20161302158441} studied the alpha-1 (8-10 Hz) and alpha-2 (10-12 Hz) waves as biomarkers to discriminate depressed and healthy subjects. The authors collected EEG data from 4 depressed and 4 control subjects using an NCC Medical data acquisition device (NCC Medical Co. Ltd., China) with 32 channels. In this research, mean absolute spectral values were extracted from alpha1 (8-10 Hz) and alpha2 (10-12 Hz) bands in two conditions of eyes closed and eyes open. The Paired T-Test statistical analysis showed lower-alpha (8-10Hz) waves in the depressed group in T5, T6, O1, O2, P3, and P4 electrodes which are located on temporal, occipital and parietal lobes.
K. Kalev et al. \cite{Kalev2015} examined Multiscale Lempel Ziv Complexity (MLZC) to exceed the accuracy of traditional LZC in discriminating depressed and control subjects. In their research, the authors collected data from 11 depressed and 11 control subjects using the Neuroscan Synamps2 (Compumedics, NC, USA) data acquisition system at the sampling frequency of 1000 Hz, which was later downsampled to 200Hz. To differentiate depressed and control subjects, the authors used the Wilcoxon Rank-Sum statistical test. The classification accuracy in LZC and MLZC was calculated with LDA and is shown in Table \ref{tab:ClassAcc}.  The highest classification accuracy was achieved with channel F3 and It was concluded that MLZC could significantly increase the classification accuracy compared to traditional LZC.

\begin{table}[htp]
\caption{\label{tab:ClassAcc} Classification accuracy of LZC and MLZC from \cite{Kalev2015}}
\centering
\resizebox{0.45\textwidth}{!}{\begin{tabular}{c}
\begin{tabular}{lll}
\hline
Channel & Classification Accuracy & Corresponding Frequency \\\hline
F3  &  86.36\%	  &  9Hz and 15.5Hz       \\
\hline
F7  &  81.82\%  &  28.5Hz and 40Hz       \\
\hline
Fp1  & 77.27\%	 &  2-6Hz, 9Hz, 10Hz and 12Hz       \\
\hline
\end{tabular}
\end{tabular}}
\end{table}

S. Akdemir Akar et al. \cite{15605052} evaluated the nonlinear analysis of EEG to differentiate depressed and control subjects. The authors used a BrainAmp DC acquisition system (Brain products GmbH, Germany) at the sampling frequency of 250Hz to collect EEG data from 15 depressed and 15 healthy subjects. The electrodes were located on prefrontal, frontal, central, parietal and temporal lobes and the ocular artifacts were removed both by data acquisition system and by visual inspection. Higuchi’s fractal Dimension (HFD), Shannon entropy, Lempel-Ziv complexity, Katz’s fractal dimension (KFD), and Kolmogorov Complexity features were extracted. The Analysis of variance (ANOVA) showed LZC, HFD, and KFD values can discriminate depressed and control subjects with higher accuracy compared to Kolmogorov complexity (KC) and Shannon entropy (ShEn) values. 
In another study by S. A. Akar et al. \cite{20160201780626}, the authors studied the wavelet analysis combined with KFD and HFD to discriminate depressed and healthy patients. The EEG data were collected from 16 depressed and 15 healthy subjects with a Brain-Amp DC acquisition system (Brain products GmbH, Germany) at 250Hz sampling frequency. The authors used Fp1, Fp2, F3, F4, F7, F8, P3, P4, P7, and P8 electrodes which are located on prefrontal, frontal and parietal lobes. Artifactes were removed by visual inspection. Fractality analysis and ANOVA statistical analysis shows that KFD and HFD values can be used to distinguish patients and healthy subjects.
A. Frantzidis et al. \cite{Frantzidis2012} explored a method with a DWT and Mahalanobis Distance (MD) classifier. A group of 33 depressed and 33 control subjects participated in their study.  To collect EEG data, they used the Nihon Kohden JE-207A (NIHON KOHDEN EUROPE GmbH) device with 57 electrodes. Signal artifacts were removed by ICA and visual inspection. Their results showed significant variation in the delta band (0.5-4 Hz) between depressed and healthy groups. 
Z. Wan et al. \cite{Wan2017} proposed an algorithm to classify patients based on EEG and Patient’s Clinical self-rating data. The self-rating emotional data used in the study is collected for two weeks by patients using an electronic self-rating scale called Profile of Mood Status and Brief Chinese Norm (POMS-BCN). The authors used a B3 EEG band (Neuro-Bridge, Japan) with a NeuroSky EEG sensor (NeuroSky, Inc.) at the sampling frequency of 512Hz to collect EEG data from 5 patients. First, Principle Component (FPC) curve was calculated based on PCA from clinical self-rating data collected by users. The signal was denoised and decomposed with DWT, and a feature vector was created with 256 features extracted from 5 EEG bands. The authors used the RF algorithm for further analysis and built a quantization model. The results showed that the proposed algorithm could compute mood statues with a high correlation with self-rating values. 
A. Suhhova et al. \cite{Suhhova2013} explored individual and traditional fixed EEG frequency bands using the Spectral Asymmetry Index (SASI) on two groups of 18 depressed and 18 control subjects. They collected EEG data from 19 electrode Cadwell Easy II (Cadwell Industries Inc., USA) data acquisition system at the sampling frequency of 400 Hz. The authors chose parietal P3 electrode for analysis. Power spectral density and EEG signal power features were extracted from the EEG data. Their results show that individual SASI analysis is more consistent than traditional fixed SASI to differentiate depressed and healthy subjects.
M. Bachmann et al. \cite{Bachmann2017} investigated single-channel EEG analysis based on DFA and SASI to discriminate depressed and healthy subjects. The authors used an 18 channel Cadwell Easy II (Cadwell Industries Inc., USA) EEG data acquisition system at the sampling frequency of 400 Hz to collect data from 17 depressed and 17 control subjects. LDA was employed to classify two groups and compute accuracy, specificity, and sensitivity. The highest performance metrics was achieved with combined SASI and DFA algorithms. It was also concluded that Pz electrode which is at Parietal lobe has the highest accuracy in discriminating two groups. 
W. Mumtaz et al. \cite{Mumtaz2015-NO2} used DFA and LR with 10-fold cross-validation to discriminate depressed and healthy subjects. In their study, EEG data were collected from 33 depressed and 33 control subjects using 19 of 24 channels with an EEG data acquisition device with BrainMaster 24 E amplifier (BrainMaster technologies Inc., USA) at the sampling frequency of 256Hz. The electrodes were placed on prefrontal, frontal, central, temporal, parietal and occipital lobes. The authors used DFA combined with three approaches to extract and select features. T-test, Wilcoxon, and Receiver Operating Characteristics (ROC) were used to rank features. The algorithm was also used with three EEG referencing methods: IR, AR and LE. The highest performance was achieved by AR with t-test. The authors concluded that the proposed algorithm could be used to discriminate healthy and depressed patients.
M. Bachmann et al. \cite{Bachmann2018} explored different methods for single channel depression diagnosis. The EEG data was collected from 13 depressed and 13 healthy participants with 30 electrodes based on International 10/20 extended system. A Neuroscan Synamps2 (Compumedics, NC, USA) was used for EEG data acquisition at the sampling frequency of 1000 Hz. The signal was visually inspected, and artifacts were removed. A combination of linear features of Alpha power variability, Relative gamma power and Spectral asymmetry index were extracted from the signal. In addition, nonlinear features of Detrended fluctuation, HFD and Lempel-ziv complexity were extracted from the signal. To measure the difference between two groups of depressed and healthy subjects, the authors used Mann-Whitney statistical test. LR classifier with leave-one-out cross-validation was also employed to classify data. The authors achieved the highest classification accuracy with the combination of linear and nonlinear features from a single channel located on central lobe.
In another study by W. Mumtaz et al. \cite{Mumtaz2016}, The authors proposed a machine learning framework to discriminate depressed and healthy subjects based on features extracted from Event-related potentials (ERP). 29 depressed and 15 healthy subjects were participated in the study and EEG data was collected with 19 wet electrodes using BrainMaster Discovery amplifier (BrainMaster technologies Inc., USA) at 256 Hz sampling frequency. Ocular and muscle artifacts were removed by Surrogate filtering method with BESA. ERP features were extracted the signal and a large dataset was created. To reduce the number of irrelevant features, three feature selection criteria of T-test, Wilcoxon and ROC were employed. The authors achieved the highest classification accuracy with Fz and with Pz located on central regions of brain. It was concluded that it is feasible to discriminate depressed and healthy subjects with single channel EEG analysis using ERP features. 
In addition to the selected papers summarized in the above, other papers, including \cite{Orgo2017-NO2}, \cite{Li2016}, \cite{Zhao2013}, \cite{Suhhova2009}, \cite{Pockberger1985}, \cite{Lee2007}, \cite{Li2007}, \cite{Li2015}, \cite{Wan2017AQA}, \cite{Hinrikus2009}, \cite{20190906547554}, and \cite{18853691} have also used statistical analysis in their research.
 In addition, functional connectivity and coherence have been used by \cite{Peng2019} , \cite{20183605764415}, \cite{17051427}, \cite{Li2016-NO2}, and \cite{20160801982396}. Also, Frontal Brain Asymmetry (FBA) which is the measure of differences between alpha power activity in right and left of the brain, was used by authors in \cite{20070710426068}. In addition,  test of variables of attention (T.O.V.A) which is a  computer-based visual attention test in clinics, was used in \cite{Ku2012}. 

\subsubsection{Information Fusion beyond EEG }

One of the critical questions for conducting EEG-based diagnosis is which cortical region of the brain to study. Several papers reported various observations about the discriminative power of various locations. In addition, several studies fuse EEG with other modalities and sometimes subjective measures to improve diagnosis. In this regard, a category of papers (such as \cite{Wan2017} and \cite{Wan2017AQA}) adopted questionnaires and self-rating tools to increase the performance of their algorithm. In addition, audio and visual stimulation techniques were also used in some studies during recording EEG signals \cite{8101151}, \cite{15605052}, \cite{Cai2017NO2}.
The first study to adopt audio stimulation was published by I. Kalatzis et al. \cite{8101151}. The authors presented the subjects with low- and high-frequency sound and a number. The subjects were asked to memorize the numbers and recall them later. S. Akdemir Akar et al. \cite{15605052} used a combination of instrumental music and sounds as a stimulus in their experiment. In another study by H. Cai et al. \cite{Cai2017NO2} the authors used an International Affective Digital Sounds (IADS) data set during EEG data acquisition. V. Ritu et al. \cite{Ritu2017} used a combination of audio and visual stimulation to study memory processing and impairment in depressed patients. The authors in \cite{Cai2018ADetection}, \cite{Zhao2017} and \cite{Cai2016PervasiveCollector} also used IADS-2 during data acquisition to further discriminate depressed patients from healthy subjects. H. Cai et al. \cite{Cai2018ACASE} used international affective digitalized sounds (IADS-2) audio stimulation while collecting EEG data to create a case-based reasoning model to diagnose depression. 
L. Ku et al. \cite{Ku2012} employed a T.O.V.A attention test during recording EEG signals. Subjects were asked to read a center cross during data acquisition. Later in another paper by Q. Zhao et al. \cite{Zhao2013} cue-target paradigm with facial expressions was studied. The subjects were presented with one of sad, happy, neutral facial expressions or object pictures during EEG recording. The authors in \cite{Mumtaz2016} used visual 3-stimulus oddball tasks as a stimulus during data acquisition.
In a study by X. Li et al. \cite{Li2015}, the authors recorded EEG signals and eye movements during a facial expression viewing task. The Chinese facial affective picture system (CFAPS) dataset was used as a stimulus during the recording. The same dataset was employed in other studies by the researchers in \cite{15798207}, \cite{XiaoweiLi2016EEG-basedClassifiers} and \cite{Bascil2016}. The authors in [37] and \cite{Wan2017AQA}  used a self-rating scale called Profile of Mood Status and Brief Chinese Norm (POMS-BCN). Subjects in \cite{Wan2017} had access to a smartphone application to rate adjectives and emotional factors. In another study by Z. Wan et al. \cite{18853691} an application to log patients’ mood status was developed. The combination of a quantitative log for mental state (Q-Log) and EEG was used to discriminate depressed patients.  


}
\end{sidewaystable*}


\begin{sidewaystable*}[htp]
\caption{\label{tab:Table2-12} Continued - summary of the most recent literature based on data processing pipeline, performance metrics, and main results}
\resizebox{1\textwidth}{!}{%
}
\end{sidewaystable*}


\begin{sidewaystable*}[htp]
\caption{\label{tab:Table2-13} Continued - summary of the most recent literature based on data processing pipeline, performance metrics, and main results}
\resizebox{1\textwidth}{!}{%
}
\end{sidewaystable*}


\begin{sidewaystable*}[htp]
\caption{\label{tab:Table2-14} Continued - summary of the most recent literature based on data processing pipeline, performance metrics, and main results}
\resizebox{1\textwidth}{!}{%
}
\end{sidewaystable*}


\begin{sidewaystable*}[htp]
\caption{\label{tab:Table2-15} Continued - summary of the most recent literature based on data processing pipeline, performance metrics, and main results}
\resizebox{1\textwidth}{!}{%
}
\end{sidewaystable*}


\begin{sidewaystable*}[htp]
\caption{\label{tab:Table2-16} Continued - summary of the most recent literature based on data processing pipeline, performance metrics, and main results}
\resizebox{1\textwidth}{!}{%
}
\end{sidewaystable*}


\begin{sidewaystable*}[htp]
\caption{\label{tab:Table2-17} Continued - summary of the most recent literature based on data processing pipeline, performance metrics, and main results}
\resizebox{1\textwidth}{!}{%

\end{tabular}}
\end{sidewaystable*}

\section{Discussion and conclusion}
This section provides an updated and comprehensive systematic review focused on depression diagnosis algorithms based on EEG signals using machine learning and statistical analyses. The literature review is motivated by the great potential that EEG has shown as an effective modality to analyze brain neurophysiological and cognitive state of patients. Considering the current state of research and the existing evidence of the richness of the information content of EEG for diagnosis of depression achieved using advanced machine learning techniques, it is imperative to promote comparative studies in a systematic manner in order to plan a convergent path which can result in the widespread use of the technology in clinics, maximizing accessibility to a depression diagnosis tool in remote areas and ultimately enabling individuals in-home monitoring of high-risk patients. 
Despite the accelerated progress of the literature, the disparity in protocols, technologies, and processing algorithms has resulted in a heterogeneous understanding of how EEG activity can be correlated with various degrees of depression in patients. Testing the performance of existing techniques on similar datasets or on datasets collected under the same data acquisition conditions and protocols may allow conducting a comparative study between the existing methods and models to find common ground. So far, this has not been achieved, resulting in scattered and limited outcomes with respect to the use of EEG for depression detection, which may not even be in perfect agreement. Some recent efforts suggest the use of public datasets \cite{18768750}, \cite{19259295}, \cite{Ke2019}, \cite{20183405723339}, \cite{Acharya2018AutomatedNetwork}, \cite{Katyal2014} to compare the benefit of various information processing pipelines for depression monitoring. For this, an available dataset for EEG-based diagnoses is the Patient Repository for EEG Data + Computational Tools (PRED+CT) \cite{Cavanagh2017}.
To address the existing scattered knowledge on how to translate EEG to depression diagnosis, a suggestion could be to propose a collective data collection to form a 'universal' anonymized EEG dataset for depression diagnosis to design a common testbed for analyzing existing computational models. This is a controversial topic, but if successful it can address a major deficit of the field, which is the disparity of the collected datasets and protocols, avoiding a convergent path to translate the technology, algorithms, and protocols into clinically viable solutions. 

In addition to the above, one of the other problems in the literature is that most of the existing techniques try to conduct a binary classification between depressed and non-depressed groups of individuals. Although this can shed light on the cortical information context related to depression, a suggestion to further improve the current state would be to collect finer clinical labels of depression as part of the data collection, instead of having only binary labels. This would require a major effort of the community but can result in a revolutionary solution for early diagnosis of depression using wearable technologies, which may be accessible in homes of people. 
Regarding the conducted literature review, we have observed variability in terms of the channels used for diagnosis, in addition to the type of information preprocessing and processing. For example, some of the studies in our systematic review have removed EEG signal artifacts manually in the preprocessing stage, such as \cite{Peng2019}, \cite{18983501}, \cite{Acharya2018AutomatedNetwork}, \cite{Bachmann2018}, and \cite{Orgo2017}. However, manual removal or subjective visual inspection of artifacts is not feasible when developing a depression diagnosis algorithm as a widely used medical device. One suggestion is to focus on neural network algorithms such as CNN that do not require preprocessing and can use raw EEG data for the conduction of the diagnosis. This trend has been observed, as reported in this paper, in the last five years, thanks to the advent of the modern machine learning algorithm and strong processing systems and cloud-based processing modules. 
In general, SVM, KNN, and CNN were used as the most frequent algorithms for the classification of patient groups. Regarding statistical analysis, the Mann-Whitney statistical test, t-test, ANOVA, and Wilcoxon Rank-Sum statistical test were mostly used in earlier studies. 
Most of the studies that do not rely on autonomous feature extraction utilized linear and nonlinear features. Among the linear feature, absolute and relative band powers, mean, max, and centroid frequency were the most used items. In addition to linear features, nonlinear features were also used for this task in the literature, among which are C0-complexity, Lempel-Ziv complexity, Shannon entropy, correlation dimension, wavelet entropy. 
Although with the use of the neural network, there is no need for feature extraction \cite{8981929}, \cite{18768750}, \cite{19259295}, \cite{Ke2019}, \cite{Sandheep2019}; however, a large dataset is required to achieve high performance. This can be addressed by increasing the number of subjects and the duration of data acquisition.   The number of EEG electrodes is another imperative factor that can significantly affect the complexity and efficacy of developed algorithms as discussed in this review article. 
A future direction for this research is to implement unobtrusive wearable systems that can be used in several clinics and even at home under the umbrella of telemedicine, which has attracted a great deal of interest lately due to the restriction in place for controlling the spread of COVID19 virus affecting the delivery of care and monitoring for other health conditions.  Thus, developing wearable systems with few EEG electrodes for depression monitoring is of high importance. There are few recent studies which uses less than 5 electrodes, including \cite{19203658}, \cite{Bachmann2018}, \cite{Bachmann2017}, and \cite{Suhhova2013}. In particular, \cite{18768750}, \cite{19259295}, \cite{18853691}, \cite{Wan2017AQA} and \cite{Wan2017} have shown the high potential of two prefrontal electrodes ( Fp1 and Fp2) to discriminate depressed patients and healthy subjects. This shows high promise for developing wearable systems with a minimal number of electrodes for depression monitoring. In this regard, finding the most informative brain regions is another imperative subject to study. After years of relevant research, there is still no agreement about the most responsible regions of the human brain involved in depression. Thus, a systematic EEG-based study on the role of different cortical regions and the corresponding changes in their activity patterns early in depression can be very important and can help with early diagnosis of depression. 
To summarize, it can be mentioned that the future direction of this field can be focused on the use of EEG for depression diagnosis as part of digital health and using cloud-based processing and IoT technologies which can allow for conducting a diagnosis of mental health conditions using widely available devices even in remote areas and ultimately in homes of patients. An ideal depression monitoring solution can be inspired by how in-home blood sugar tests can now be found in homes of patients allowing for better monitoring and management. Thus, extensive research into the design and development of cloud-based algorithms and platforms for IoT-based depression diagnosis still has to be conducted to develop lower complexity algorithms to extract, select, and encrypt vital features and transfer extracted data to a cloud-based server for further analysis (a micro-macro processing model).  
This literature review aimed to collect, categorize, and discuss the advanced research in the field of EEG-based depression monitoring to shed light on the future directions of research and development. Based on the conducted literature review, it can be mentioned that there is a wide range of EEG-based studies supporting the use of this technology for depression detection. The future of this field, as mentioned, can be focused on designing in-home and widely accessible technology and providing a finer grade of depression.

\section{APPENDIX A}
Summary of the most recent literature categorized based on dataset, experiment duration, number of subjects and their gender, age, selection criteria and condition of the experiment.
See Tables \ref{tab:Table3-11} to \ref{tab:Table3-5}. 

\section{APPENDIX B}
Summary of the most recent literature based on type and number of electrodes, their position, data acquisition device, and sampling frequency, impedance and the type of filters used.
See Tables \ref{tab:Table4-1} to \ref{tab:Table4-155}.

\section{APPENDIX C}
Types of features used in most recent literature.
See Tables \ref{tab:Table5-1} to \ref{tab:Table5-4}..
\vfill

\begin{sidewaystable*}[htp]
\caption{\label{tab:Table3-11} Summary of the most recent literature categorized based on dataset, experiment duration, number of subjects and their gender, age, selection criteria and condition of the experiment}
\resizebox{1\textwidth}{!}{\begin{tabular}{lllllllll}

\textbf{\#}           & \textbf{Own dataset}                                                                                              & \textbf{Duration (for analysis)}                                                            & \textbf{Subjects/Gender}                                                                                                                                                                  & \textbf{Age (Mean ±  SD)}                                                                                                                            & \textbf{Participant selection and rating scale}                                                                    & \textbf{Experiment condition}                                                                                                                    \\

\rowcolor[HTML]{F2F2F2} 
\cite{Nassibi2022}       & Public dataset                                                                                                    & 7 minutes                                                                                                                                                                                                                        & \begin{tabular}[c]{@{}l@{}}42 depressed(30F, 12M)\\ 42 healthy(26F, 16M)\end{tabular}                                                                                                     & \begin{tabular}[c]{@{}l@{}}Depressed: \\ F(18.46 ± 0.77), M(19.08 ± 1.67)\\ Healthy:\\ F(19 ± 1.01),  M(19.13 ± 1.4)\end{tabular}                                                      & SCID, BDI                                                                                                                                                    & Eyes closed and eyes open                                                                                                                        \\

\cite{Movahed2022}   & Public dataset                                                                                                    & 5 minutes                                                                                   & \begin{tabular}[c]{@{}l@{}}34 MDD (17F,   17M)\\ 30   healthy (9F, 21M)\end{tabular}                                                                                                      & \begin{tabular}[c]{@{}l@{}}MDD: 40.3 ± 12.9) \\ Healthy: 38.3 ± 15.6\end{tabular}                                                                    & DSM-IV                                                                                                             & Eyes closed and eyes open                                                                                                                        \\
\rowcolor[HTML]{F2F2F2} 
\cite{Wang2DCNN2022} & \begin{tabular}[c]{@{}l@{}}Provided by 3A \\ grade hospital\end{tabular}                                          & 2 to 4 minutes                                                                              & \begin{tabular}[c]{@{}l@{}}16 depressed\\ 16 healthy\end{tabular}                                                                                                                         & Did not mention                                                                                                                                      & Did not mention                                                                                                    & Resting-state (eyes closed)                                                                                                                      \\
\cite{Li2022}        & Own dataset                                                                                                       & 30 seconds                                                                                  & 92 depressed                                                                                                                                                                              & Did not mention                                                                                                                                      & HAMD-17                                                                                                            & Resting-state (eyes closed)                                                                                                                      \\
\rowcolor[HTML]{F2F2F2} 
\cite{Garg2022}      & Public EEG dataset                                                                                                & 5 minutes                                                                                   & \begin{tabular}[c]{@{}l@{}}62 depressed, \\ 58 healthy\end{tabular}                                                                                                                       & 12 to 77 years                                                                                                                                       & Did not mention                                                                                                    & Resting-state (eyes closed)                                                                                                                      \\
\cite{Kabbara2022}   & \begin{tabular}[c]{@{}l@{}}Public EEG dataset \\ and own dataset\end{tabular}                                     & \begin{tabular}[c]{@{}l@{}}Dataset 1 and 2: 5 minutes,\\ Dataset 3: 16 minutes\end{tabular} & \begin{tabular}[c]{@{}l@{}}Dataset 1: 45 Depressed, \\ 76 healthy\\ Dataset 2: 24 depressed \\ (11F, 13M), \\ 29 healthy (9F, 20M)\\ Dataset 3: 154 healthy \\ (45F,   109M)\end{tabular} & \begin{tabular}[c]{@{}l@{}}Dataset 1: 18–25\\ Dataset 2: \\ depressed (30.88 ± 10.37)\\ healthy (31.45 ± 9.15) \\ Dataset 3: 25.1 ± 3.1\end{tabular} & \begin{tabular}[c]{@{}l@{}}Dataset 1: eMINI\\ Dataset 2: MINI, DSM, HDRS\\ Dataset 3: did not mention\end{tabular} & \begin{tabular}[c]{@{}l@{}}Dataset 1: resting-state\\ Dataset 2: resting-state (eyes closed)\\ Dataset 3: eyes open (resting-state)\end{tabular} \\
\rowcolor[HTML]{F2F2F2} 
\cite{Sharmila2022}  & Public dataset                                                                                                    & Did not mention                                                                             & Did not mention                                                                                                                                                                           & Did not mention                                                                                                                                      & Did not mention                                                                                                    & Did not mention                                                                                                                                  \\
\cite{Zhao2022}      & Own dataset                                                                                                       & 4 minutes                                                                                   & \begin{tabular}[c]{@{}l@{}}40 depressed females,\\ 38 healthy females\end{tabular}                                                                                                        & \begin{tabular}[c]{@{}l@{}}Depressed: 18.51 ± 0.42\\ Healthy: 18.75 ± 0.36\end{tabular}                                                              & \begin{tabular}[c]{@{}l@{}}Examined by psychologists,\\ BDI-II, SDS\end{tabular}                                   & Resting-state (eyes closed and open)                                                                                                             \\
\rowcolor[HTML]{F2F2F2} 
\cite{Kim2022}       & \begin{tabular}[c]{@{}l@{}}Obtained from National \\ Standard Reference Data\\ Center for Korean EEG\end{tabular} & 5 minutes                                                                                   & \begin{tabular}[c]{@{}l@{}}116 (95F, 23M) depressed\\ 80 (36F, 44M) healthy\end{tabular}                                                                                                  & \begin{tabular}[c]{@{}l@{}}Depressed: 58.66 ± 15.08\\ Healthy: 48.66 ± 16.71\end{tabular}                                                            & BDI                                                                                                                & Resting-state (eyes closed)                                                                                                                      \\
\cite{Zhang2022}     & Own dataset                                                                                                       & 9-10 hours                                                                                  & \begin{tabular}[c]{@{}l@{}}30 depressed (12F, 18M), \\ 30 healthy (17F, 13M)\end{tabular}                                                                                                 & \begin{tabular}[c]{@{}l@{}}Depressed: 22.13 ± 7.07\\ Healthy: 19.67 ± 1.25\end{tabular}                                                              & \begin{tabular}[c]{@{}l@{}}Examined by psychiatrists \\ based on Diagnostic and \\ DSM-IV, HAMD, SDS\end{tabular}  & Overnight sleep                                                                                                                                  \\
\rowcolor[HTML]{F2F2F2} 
\cite{Lin2022}       & Own dataset                                                                                                       & 30 seconds                                                                                  & \begin{tabular}[c]{@{}l@{}}138 in total (91F, 47M), \\ analysis: 34   depressed, \\ 58 healthy\end{tabular}                                                                               & Total (38.78 ± 13.56)                                                                                                                                & HAMD-17                                                                                                            & Eyes closed                                                                                                                                      \\
\cite{Song2022}      & Own dataset                                                                                                       & 5.5 minutes                                                                                 & \begin{tabular}[c]{@{}l@{}}40 depressed (25F, 15M), \\ 40 healthy (23F, 17M)\end{tabular}                                                                                                 & \begin{tabular}[c]{@{}l@{}}Depressed (18 –79 (45.5 average)), \\ Healthy (22 – 73 (44.9 average))\end{tabular}                                       & ICD-10, HDRS                                                                                                       & Resting-state (eyes closed)                                                                                                                      \\
\rowcolor[HTML]{F2F2F2} 
\cite{Avots2022}     & Own dataset                                                                                                       & \begin{tabular}[c]{@{}l@{}}20 minutes in total \\ (5 minutes was used)\end{tabular}         & 10 depressed, 10 healthy                                                                                                                                                                  & 24-60 years depressed and healthy                                                                                                                    & HAM-D and EST-Q                                                                                                    & Resting-state (eyes closed)                                                                                                                      \\
\cite{Sun2022}       & Own dataset                                                                                                       & 5 minutes                                                                                   & 16 depressed, 16 healthy                                                                                                                                                                  & Did not mentioned                                                                                                                                    & \begin{tabular}[c]{@{}l@{}}MINI, DSM-IV, PHQ-9,   \\ GAD-7, PSQI\end{tabular}                                      & Resting-state (eyes closed)                                                                                                                      \\
\rowcolor[HTML]{F2F2F2} 
\cite{Wang2022}      & Two public datasets                                                                                               & did not mentioned                                                                           & \begin{tabular}[c]{@{}l@{}}Dataset 1: 18 depressed, \\ 25 Normal\\ Dataset 2: did not mentioned\end{tabular}                                                                              & \begin{tabular}[c]{@{}l@{}}Dataset 1: 18 -53\\ Dataset 2: 18-25\end{tabular}                                                                         & \begin{tabular}[c]{@{}l@{}}Dataset 1: did not mentioned,\\ Dataset 2: BDI\end{tabular}                             & \begin{tabular}[c]{@{}l@{}}Dataset 1: Resting-state (eyes closed)\\ Dataset 2: did not mentioned\end{tabular}                                    \\
\cite{Nayad2022}    & Public dataset                                                                                                    & 5 minutes                                                                                   & \begin{tabular}[c]{@{}l@{}}34 depressed, \\ 30 healthy\end{tabular}                                                                                                                       & \begin{tabular}[c]{@{}l@{}}Depressed: 40.3 ± 12.9, \\ healthy: 38.3 ± 15.6\end{tabular}                                                              & Did not mentioned                                                                                                  & \begin{tabular}[c]{@{}l@{}}Eyes open (EO), eyes closed (EC), \\ cognitive task (P300)\end{tabular}                                               \\
\rowcolor[HTML]{F2F2F2} 
\cite{Babu2022}      & Did not mentioned                                                                                                 & 30 second sessions                                                                          & \begin{tabular}[c]{@{}l@{}}Moderate depression: 29, \\ mild depression: 29, \\ healthy 30\end{tabular}                                                                                    & Did not mentioned                                                                                                                                    & BDI-II, PHQ-9                                                                                                      & Resting-state (eyes closed and open)                                                                                                             \\
\cite{Uyulan2022}    & Own dataset                                                                                                       & 3 minutes                                                                                   & \begin{tabular}[c]{@{}l@{}}24 depressed (16F, 8M), \\ 24 healthy (13F, 11M)\end{tabular}                                                                                                  & \begin{tabular}[c]{@{}l@{}}Depressed (54.2±18.0), \\ healthy (45.9±16.1)\end{tabular}                                                                & \begin{tabular}[c]{@{}l@{}}(DSM)-IV, \\ MADRS, FIGS\end{tabular}                                                   & Resting-state (eyes closed)                                                                                                                     
     
\end{tabular}}
\end{sidewaystable*}

\begin{sidewaystable*}[htp]
\caption{\label{tab:Table3-12} Continued - summary of the most recent literature categorized based on dataset, experiment duration, number of subjects and their gender, age, selection criteria and condition of the experiment}
\resizebox{1\textwidth}{!}{\begin{tabular}{lllllllll}

\textbf{\#}             & \textbf{Own dataset}                                                                 & \textbf{Duration (for analysis)}                                                                                     & \textbf{Subjects/Gender}                                                                     & \textbf{Age (Mean ±  SD)}                                                                                            & \textbf{Participant selection and rating scale}                                                                               & \textbf{Experiment condition}                                                                   \\
\rowcolor[HTML]{F2F2F2} 
\cite{Goswami2022}     & Public dataset                                                                       & Did not mentioned                                                                                                    & \begin{tabular}[c]{@{}l@{}}10 depressed, \\ 12 past depression, \\ 7 healthy\end{tabular}    & Did not mentioned                                                                                                    & BDI                                                                                                                           & Did not mentioned                                                                               \\
\cite{Jan2022}         & Own dataset                                                                          & \begin{tabular}[c]{@{}l@{}}3 minutes eyes closed, \\ 3 minutes eyes open,\end{tabular}                               & \begin{tabular}[c]{@{}l@{}}26 depressed (23F, 3M),\\ 24 healthy (14F, 10M)\end{tabular}      & \begin{tabular}[c]{@{}l@{}}Depressed (22.43 ± 1.4),\\ healthy (19.6 ± 2.5)\end{tabular}                              & BDI-II (online)                                                                                                               & Resting-state (eyes closed and open)                                                            \\
\rowcolor[HTML]{F2F2F2} 
\cite{Seal2022}        & Own dataset                                                                          & 9 minutes                                                                                                            & \begin{tabular}[c]{@{}l@{}}15 depressed, \\ 18 healthy\end{tabular}                          & 19-36 years (21.45 ± 3.29)                                                                                           & PHQ-9                                                                                                                         & Resting-state (eyes closed and open)                                                            \\
\cite{Thakare2022}     & Public dataset                                                                       & Did not mentioned                                                                                                    & Did not mentioned                                                                            & Did not mentioned                                                                                                    & Did not mentioned                                                                                                             & Did not mentioned                                                                               \\
\rowcolor[HTML]{F2F2F2} 
\cite{Ghiasi2021}      & Own dataset                                                                          & 4 minutes                                                                                                            & \begin{tabular}[c]{@{}l@{}}26 depressed (dysphoria, 26F),   \\ 38 healthy (38F)\end{tabular} & Average age of 22                                                                                                    & BDI-II, SCID-I                                                                                                                & Resting-state (eyes open)                                                                       \\
\cite{Loh2021}         & Public dataset                                                                       & \begin{tabular}[c]{@{}l@{}}5 minutes eyes closed, \\ 5 minutes eyes open, \\ 10 minutes visual stimulus\end{tabular} & 34 depressed, 30 healthy                                                                     & 12-77 years                                                                                                          & Did not mentioned                                                                                                             & \begin{tabular}[c]{@{}l@{}}Eyes open (EO), \\ eyes closed (EC), \\ visual stimulus\end{tabular} \\
\rowcolor[HTML]{F2F2F2}

\cite{Movahed2021}     & Public EEG dataset                                                                   & \begin{tabular}[c]{@{}l@{}}1 minute EEG samples, \\ (249 MDD and 261 \\ Healthy samples)\end{tabular}                & \begin{tabular}[c]{@{}l@{}}34 MDD (17F, 17M)\\ 30 healthy (9F, 21M)\end{tabular}             & \begin{tabular}[c]{@{}l@{}}MDD: 40.3 + 12.9) \\ Healthy: 38.3 + 15.6\end{tabular}                                    & DSM-IV                                                                                                                        & Eyes closed and eyes open                                                                       \\
\cite{Li2021}          & \begin{tabular}[c]{@{}l@{}}No, data acquired \\ from Lanzhou University\end{tabular} & \begin{tabular}[c]{@{}l@{}}6 seconds trials, \\ 30 pieces of data\end{tabular}                                       & \begin{tabular}[c]{@{}l@{}}10 depressed (4F, 6M)\\ 10 healthy (2F, 8M)\end{tabular}          & \begin{tabular}[c]{@{}l@{}}Depressed: 20.6 ± 1.74,   \\ healthy: 20.3 ± 1.85\end{tabular}                            & \begin{tabular}[c]{@{}l@{}}Questionnaires and scale screening, \\ no psychopathological diseases\end{tabular}                 & Eyes open                                                                                       \\
\rowcolor[HTML]{F2F2F2} 
\cite{Sharma2021}      & Public dataset                                                                       & \begin{tabular}[c]{@{}l@{}}5 minutes eyes open \\ and 5 minutes eyes closed\end{tabular}                             & \begin{tabular}[c]{@{}l@{}}34 depressed\\ 30 healthy\\ (24F, 40M)\end{tabular}               & 12 to 77 years (mean 20.54)                                                                                          & DSM-IV                                                                                                                        & Eyes closed and eyes open                                                                       \\
\cite{Savinov2021}     & Yes, data collected                                                                  & \begin{tabular}[c]{@{}l@{}}3 minutes eyes open \\ and 3 minutes eyes closed\end{tabular}                             & \begin{tabular}[c]{@{}l@{}}23 depressed, \\ 9 healthy,\\ (24F, 8M)\end{tabular}              & 41.9 ± 14.6                                                                                                          & BDI                                                                                                                           & Eyes closed and eyes open                                                                       \\
\rowcolor[HTML]{F2F2F2} 
\cite{Uyulan2021}      & Own dataset                                                                          & 200 seconds                                                                                                          & 46 depressed, 46 healthy                                                                     & \begin{tabular}[c]{@{}l@{}}Depressed: 22 to 51 years (average 39.3),\\ Healthy: 20 to 45 (average 37.1)\end{tabular} & \begin{tabular}[c]{@{}l@{}}HAM-D, \\ Composite International Diagnostic \\ Interview (CIDI): ICD-10\\ and DSM-IV\end{tabular} & Resting-state (eyes closed)                                                                     \\
\cite{Wang2021}        & MODMA public dataset                                                                 & 5 minutes                                                                                                            & \begin{tabular}[c]{@{}l@{}}24 depressed (11F, 13M), \\ 29 healthy (9F, 20M)\end{tabular}     & \begin{tabular}[c]{@{}l@{}}Depressed (30.88 ± 10.3), \\ healthy (31.45 ±   9.15)\end{tabular}                        & DSM                                                                                                                           & Resting-state                                                                                   \\
\rowcolor[HTML]{F2F2F2} 
\cite{Jiang2021}       & \begin{tabular}[c]{@{}l@{}}Own dataset + \\ Ekman emotion database\end{tabular}      & 30 minutes                                                                                                           & \begin{tabular}[c]{@{}l@{}}16 depressed (10F, 6M), \\ 14 healthy (10F, 4M)\end{tabular}      & Depressed (37.75 ± 14.19), healthy (40.86 ± 12.29)                                                                   & HAMD, SAS, SDS                                                                                                                & Viewing task (Eyes open)                                                                        \\
\cite{Hong2021}        & Own dataset                                                                          & 30 seconds                                                                                                           & \begin{tabular}[c]{@{}l@{}}34 depressed, \\ 58 healthy\end{tabular}                          & Did not mention                                                                                                      & HAMD-17                                                                                                                       & Resting-state (eyes closed)                                                                     \\
\rowcolor[HTML]{F2F2F2} 
\cite{WeiEEGBASED2021} & Own dataset                                                                          & 90 seconds                                                                                                           & \begin{tabular}[c]{@{}l@{}}81 depressed, \\ 89 healthy\end{tabular}                          & 18 - 65                                                                                                              & MINI, PHQ-9                                                                                                                   & Resting-state (eyes closed)                                                                     \\
\cite{Seal2021}        & Did not mentioned                                                                    & 9 minutes                                                                                                            & \begin{tabular}[c]{@{}l@{}}15 depressed, \\ 18 healthy\end{tabular}                          & Did not mentioned                                                                                                    & PHQ-9                                                                                                                         & Resting-state                                                                                   \\
\rowcolor[HTML]{F2F2F2} 
\cite{9477800}         & Own dataset                                                                          & 7 minutes                                                                                                            & \begin{tabular}[c]{@{}l@{}}16 depressed, \\ 16 healthy\end{tabular}                          & Did not mentioned                                                                                                    & Did not mention                                                                                                               & Resting-state (eyes closed)                                                                     \\
\cite{Shen2021}        & Public dataset (MODMA)                                                               & 5 minutes                                                                                                            & \begin{tabular}[c]{@{}l@{}}15 depressed, \\ 20 healthy\end{tabular}                          & 18 - 55                                                                                                              & LES, MINI, CTQ                                                                                                                & Resting-state (eyes closed)                                                                    
\end{tabular}}
\end{sidewaystable*}

\begin{sidewaystable*}[htp]
\caption{\label{tab:Table3-13} Continued - summary of the most recent literature categorized based on dataset, experiment duration, number of subjects and their gender, age, selection criteria and condition of the experiment}
\resizebox{1\textwidth}{!}{\begin{tabular}{lllllllll}

\textbf{\#}               & \textbf{Own dataset}                                                                                    & \textbf{Duration (for analysis)}                                                                                                                                                                                                 & \textbf{Subjects/Gender}                                                                                                                                                           & \textbf{Age (Mean ±  SD)}                                                                                                                                                                 & \textbf{Participant selection and rating scale}                                                             & \textbf{Experiment condition}                                                                       \\
\cite{Duan2021}          & Own dataset                                                                                             & Did not mentioned                                                                                                                                                                                                                & 35 unipolar, 21 bipolar, 35 healthy                                                                                                                                                & Did not mentioned                                                                                                                                                                         & Did not mentioned                                                                                           & Resting-state                                                                                       \\
\rowcolor[HTML]{F2F2F2} 
\cite{Zhu2020}           & \begin{tabular}[c]{@{}l@{}}Yes, data collected by \\ Lanzhou University \\ Second Hospital\end{tabular} & Did not mention                                                                                                                                                                                                                  & \begin{tabular}[c]{@{}l@{}}Eye tracking dataset: \\ 18 depressed (9F, 9M), \\ 18 healthy (8F, 10M)\\ EEG dataset: \\ 17 depressed (8F, 9M), \\ 17   healthy (7F, 10M)\end{tabular} & \begin{tabular}[c]{@{}l@{}}Eye tracking dataset: \\ (Depressed: 31.56 ± 8.6\\ \\ Healthy: 31.33 ± 9.26)\\ EEG dataset: \\ (Depressed: 29.32 ± 89.23\\ Healthy: 30.61 ± 9.44)\end{tabular} & PHQ-9                                                                                                       & \begin{tabular}[c]{@{}l@{}}Eyes open (eye tracking) \\ and Eyes closed (resting-state)\end{tabular} \\
\cite{Saeedi2020}        & Public dataset                                                                                          & 5 minutes                                                                                                                                                                                                                        & \begin{tabular}[c]{@{}l@{}}34 depressed (17F, 17M), \\ 30 healthy (9F, 21M)\end{tabular}                                                                                           & \begin{tabular}[c]{@{}l@{}}Depressed (27 to 53 years, \\ Healthy (22 to 5 years)\end{tabular}                                                                                             & Did not mention                                                                                             & Eyes closed                                                                                         \\
\rowcolor[HTML]{F2F2F2} 
\cite{Duan2020}          & Own dataset                                                                                             & 3 minutes                                                                                                                                                                                                                        & 16 depressed, 16 healthy                                                                                                                                                           & Did not mention                                                                                                                                                                           & DSM-IV                                                                                                      & Relaxed (awake)                                                                                     \\
\cite{Apsari2020}        & Two public datasets                                                                                     & \begin{tabular}[c]{@{}l@{}}First dataset: \\ 5 minutes eyes closed, \\ 5 minutes eyes open, \\ 10 minutes 3-stimulus \\ visual oddball task, \\ Second dataset: \\ 8 minutes eyes closed \\ and 8 minutes eyes open\end{tabular} & \begin{tabular}[c]{@{}l@{}}First dataset: 28 depressed, 27 healthy\\ Second dataset: 216 subjects\end{tabular}                                                                     & Did not mention                                                                                                                                                                           & HAMD                                                                                                        & \begin{tabular}[c]{@{}l@{}}Resting-state \\ (eyes closed and open)\end{tabular}                     \\
\rowcolor[HTML]{F2F2F2} 
\cite{Saleque2020}       & Public dataset                                                                                          & \begin{tabular}[c]{@{}l@{}}5 minutes eyes open and \\ 5 minutes eyes closed\end{tabular}                                                                                                                                         & \begin{tabular}[c]{@{}l@{}}34 (17F, 17M) depressed\\ 30 (9F, 21M) healthy\end{tabular}                                                                                             & Did not mention                                                                                                                                                                           & Did not mention                                                                                             & Eyes closed and eyes open                                                                           \\
\cite{Sun2020}           & Own dataset                                                                                             & 5 minutes                                                                                                                                                                                                                        & \begin{tabular}[c]{@{}l@{}}24 depressed (11F, 13M), \\ 29 healthy (9F, 20M)\end{tabular}                                                                                           & \begin{tabular}[c]{@{}l@{}}Depressed (30.88 ± 10.37), \\ healthy (31.45 ± 9.15)\end{tabular}                                                                                              & \begin{tabular}[c]{@{}l@{}}Diagnosed by psychiatrists, \\ MINI, DSM-IV, PHQ-9\end{tabular}                  & Resting-state (eyes closed)                                                                         \\
\rowcolor[HTML]{F2F2F2} 
\cite{Thoduparambil2020} & Public dataset                                                                                          & Did not mentioned                                                                                                                                                                                                                & Did not mentioned                                                                                                                                                                  & Did not mentioned                                                                                                                                                                         & Did not mentioned                                                                                           & Eyes open                                                                                           \\
\cite{Mahato2020}        & Own dataset                                                                                             & 15 minutes                                                                                                                                                                                                                       & 24 depressed, 20 healthy                                                                                                                                                           & \begin{tabular}[c]{@{}l@{}}Depressed: 35 ± 5.9, \\ healthy: 36 ± 4.2\end{tabular}                                                                                                         & DSM-V, HAM-D                                                                                                & Resting-state (eyes closed)                                                                         \\
\rowcolor[HTML]{F2F2F2} 
\cite{Mahato2020_2}     & Public dataset                                                                                          & 5 minutes                                                                                                                                                                                                                        & \begin{tabular}[c]{@{}l@{}}34 depressed (17F, 17M), \\ 30 healthy (9F, 21M), \\ only 30 depressed and \\ 30 healthy subjects were used\end{tabular}                                & \begin{tabular}[c]{@{}l@{}}Depressed (40.3 ± 12.9), \\ healthy (38.3 ± 15.6)\end{tabular}                                                                                                 & DSM IV                                                                                                      & Resting-state (eyes closed)                                                                         \\
\cite{Kang2020}          & Public dataset                                                                                          & \begin{tabular}[c]{@{}l@{}}5 minutes eyes closed, \\ 5 minutes eyes open, \\ 10 minutes visual stimulus\end{tabular}                                                                                                             & 34 depressed, 30 healthy                                                                                                                                                           & \begin{tabular}[c]{@{}l@{}}Depressed (40.33±12.861), \\ healthy (38.227± 15.64)\end{tabular}                                                                                              & DSM-IV                                                                                                      & \begin{tabular}[c]{@{}l@{}}Eyes open (EO), \\ eyes closed (EC), \\ visual stimulus\end{tabular}     \\
\rowcolor[HTML]{F2F2F2} 
\cite{Ding2019}          & Own dataset                                                                                             & \begin{tabular}[c]{@{}l@{}}20 trails each lasted \\ for 10 seconds\end{tabular}                                                                                                                                                  & \begin{tabular}[c]{@{}l@{}}144 depressed (86F, 58M), \\ 204 healthy (136F, 68M)\end{tabular}                                                                                       & \begin{tabular}[c]{@{}l@{}}Depressed (27.65 ± 9.5), \\ healthy (27.46 ±   9.61)\end{tabular}                                                                                              & \begin{tabular}[c]{@{}l@{}}Diagnosis by psychiatrists and \\ SDS (Zung et al.,   1965)\end{tabular}         & Eyes open (EO)                                                                                      \\
\cite{Cai2018StudyFS}    & Own dataset                                                                                             & 30 minutes                                                                                                                                                                                                                       & \begin{tabular}[c]{@{}l@{}}152 depressed,    \\ 113 healthy\end{tabular}                                                                                                           & 18 - 55                                                                                                                                                                                   & \begin{tabular}[c]{@{}l@{}}Examined by psychologists and \\ various rating scales (no details)\end{tabular} & Resting-state (eyes closed)                                                                         \\
\rowcolor[HTML]{F2F2F2} 
\cite{Li2015Second}      & Own dataset                                                                                             & \begin{tabular}[c]{@{}l@{}}180 seconds (30 segments, \\ each 6 second)\end{tabular}                                                                                                                                              & \begin{tabular}[c]{@{}l@{}}In total 36 subjects (12F, 24M), \\ for   analysis: 9 depressed (4F, 5M), \\ 9 healthy (did not mentioned the genders)\end{tabular}                     & Did not mention                                                                                                                                                                           & BDI-II                                                                                                      & Viewing task (Eyes open)                                                                           
\end{tabular}}
\end{sidewaystable*}

\begin{sidewaystable*}[htp]
\caption{\label{tab:Table3-1} Continued - summary of the most recent literature categorized based on dataset, experiment duration, number of subjects and their gender, age, selection criteria and condition of the experiment}
\resizebox{1\textwidth}{!}{\begin{tabular}{lllllllll}

\multicolumn{1}{c}{}                                       & \multicolumn{1}{c}{}                                                & \multicolumn{1}{c}{}                                                            & \multicolumn{1}{c}{}                                                                                                                            &                                                                                                                                                        & \multicolumn{1}{c}{}                                                                                                                                       & \multicolumn{1}{c}{}                                                                            \\
\multicolumn{1}{c}{\multirow{-2}{*}{\textit{\textbf{\#}}}} & \multicolumn{1}{c}{\multirow{-2}{*}{\textit{\textbf{Own dataset}}}} & \multicolumn{1}{c}{\multirow{-2}{*}{\textit{\textbf{Duration (for analysis)}}}} & \multicolumn{1}{c}{\multirow{-2}{*}{\textit{\textbf{Subjects/Gender}}}}                                                                         & \multirow{-2}{*}{\textit{\textbf{Age (Mean ± SD)}}}                                                                                                   & \multicolumn{1}{c}{\multirow{-2}{*}{\textit{\textbf{Participant selection and rating scale}}}}                                                             & \multicolumn{1}{c}{\multirow{-2}{*}{\textit{\textbf{Experiment condition}}}}                    \\
\rowcolor[HTML]{F2F2F2} 
\cellcolor[HTML]{F2F2F2}                                   & \cellcolor[HTML]{F2F2F2}                                            & \cellcolor[HTML]{F2F2F2}                                                        & 12 (6F, 6M) unmedicated                                                                                                                         & Unmedicated: 28.6±7.3                                                                                                                                & \cellcolor[HTML]{F2F2F2}                                                                                                                                   & \cellcolor[HTML]{F2F2F2}                                                                        \\
\rowcolor[HTML]{F2F2F2} 
\cellcolor[HTML]{F2F2F2}                                   & \cellcolor[HTML]{F2F2F2}                                            & \cellcolor[HTML]{F2F2F2}                                                        & 11 (6F, 5M) medicated                                                                                                                           & Medicated: 29.8±10.6                                                                                                                                 & \cellcolor[HTML]{F2F2F2}                                                                                                                                   & \cellcolor[HTML]{F2F2F2}                                                                        \\
\rowcolor[HTML]{F2F2F2} 
\multirow{-3}{*}{\cellcolor[HTML]{F2F2F2}\cite{8981929}}         & \multirow{-3}{*}{\cellcolor[HTML]{F2F2F2}Yes}                       & \multirow{-3}{*}{\cellcolor[HTML]{F2F2F2}5 minutes}                             & 12 (6F, 6M) healthy                                                                                                                             & Healthy: 26.4±9.8                                                                                                                                    & \multirow{-3}{*}{\cellcolor[HTML]{F2F2F2}Clinical interview by a clinician}                                                                                & \multirow{-3}{*}{\cellcolor[HTML]{F2F2F2}Resting-state (eyes closed)}                           \\
                                                           &                                                                     &                                                                                 & 34 (17F, 17M) depressed                                                                                                                         & Depressed: 40.3±12.9                                                                                                                                 &                                                                                                                                                            &                                                                                                 \\
\multirow{-2}{*}{\cite{18768750}}                                 & \multirow{-2}{*}{No}                                                & \multirow{-2}{*}{5 minutes}                                                     & 30 (9F, 21M) healthy                                                                                                                            & Healthy: 38.3±15.6                                                                                                                                   & \multirow{-2}{*}{DSM-IV, BDI-II, HADS}                                                                                                                     & \multirow{-2}{*}{Resting-state (eyes closed)}                                                   \\
\rowcolor[HTML]{F2F2F2} 
\cellcolor[HTML]{F2F2F2}                                   & \cellcolor[HTML]{F2F2F2}                                            & \cellcolor[HTML]{F2F2F2}                                                        & 14 depressed                                                                                                                                    & \cellcolor[HTML]{F2F2F2}                                                                                                                               & \cellcolor[HTML]{F2F2F2}                                                                                                                                   & \cellcolor[HTML]{F2F2F2}                                                                        \\
\rowcolor[HTML]{F2F2F2} 
\multirow{-2}{*}{\cellcolor[HTML]{F2F2F2}\cite{20193307311994}}         & \multirow{-2}{*}{\cellcolor[HTML]{F2F2F2}Yes}                       & \multirow{-2}{*}{\cellcolor[HTML]{F2F2F2}19 x 8 seconds trials}                 & 14   healthy                                                                                                                                    & \multirow{-2}{*}{\cellcolor[HTML]{F2F2F2}Did not mention}                                                                                              & \multirow{-2}{*}{\cellcolor[HTML]{F2F2F2}MINI, DSM-IV}                                                                                                     & \multirow{-2}{*}{\cellcolor[HTML]{F2F2F2}Eyes open with visual stimulation}                     \\
                                                           &                                                                     &                                                                                 & 34 (17F, 17M) depressed                                                                                                                         & Depressed: 40.3±12.9                                                                                                                                 &                                                                                                                                                            &                                                                                                 \\
\multirow{-2}{*}{\cite{19259295}}                                 & \multirow{-2}{*}{No}                                                & \multirow{-2}{*}{5 minutes}                                                     & 30 (9F, 21M) healthy                                                                                                                            & Healthy: 38.3±15.6                                                                                                                                   & \multirow{-2}{*}{DSM-IV, BDI-II, HADS}                                                                                                                     & \multirow{-2}{*}{Resting-state (eyes closed)}                                                   \\
\rowcolor[HTML]{F2F2F2} 
\cellcolor[HTML]{F2F2F2}                                   & \cellcolor[HTML]{F2F2F2}                                            & \cellcolor[HTML]{F2F2F2}                                                        & \cellcolor[HTML]{F2F2F2}                                                                                                                        & \cellcolor[HTML]{F2F2F2}                                                                                                                               & \cellcolor[HTML]{F2F2F2}                                                                                                                                   & \cellcolor[HTML]{F2F2F2}                                                                        \\
\rowcolor[HTML]{F2F2F2} 
\cellcolor[HTML]{F2F2F2}                                   & \cellcolor[HTML]{F2F2F2}                                            & \cellcolor[HTML]{F2F2F2}                                                        & \cellcolor[HTML]{F2F2F2}                                                                                                                        & \cellcolor[HTML]{F2F2F2}                                                                                                                               & \cellcolor[HTML]{F2F2F2}                                                                                                                                   & \cellcolor[HTML]{F2F2F2}                                                                        \\
\rowcolor[HTML]{F2F2F2} 
\cellcolor[HTML]{F2F2F2}                                   & \cellcolor[HTML]{F2F2F2}                                            & \cellcolor[HTML]{F2F2F2}                                                        & \multirow{-3}{*}{\cellcolor[HTML]{F2F2F2}\begin{tabular}[c]{@{}l@{}}First study:\\ 34 (17F, 17M) depressed\\ 30 (9F, 21M) healthy\end{tabular}} & \cellcolor[HTML]{F2F2F2}                                                                                                                               & \cellcolor[HTML]{F2F2F2}                                                                                                                                   & \cellcolor[HTML]{F2F2F2}                                                                        \\
\rowcolor[HTML]{F2F2F2} 
\multirow{-4}{*}{\cellcolor[HTML]{F2F2F2}\cite{Ke2019}}         & \multirow{-4}{*}{\cellcolor[HTML]{F2F2F2}No}                        & \multirow{-4}{*}{\cellcolor[HTML]{F2F2F2}Did not mention}                       & \begin{tabular}[c]{@{}l@{}}Second study: \\ 17 depressed\end{tabular}                                                                           & \multirow{-4}{*}{\cellcolor[HTML]{F2F2F2}\begin{tabular}[c]{@{}l@{}}First study: \\ (depressed: 40.3±12.9, \\ healthy: 38.227 ± 15.64)\end{tabular}} & \multirow{-4}{*}{\cellcolor[HTML]{F2F2F2}\begin{tabular}[c]{@{}l@{}}All the subjects were screened for\\ possible mental or physical illness\end{tabular}} & \multirow{-4}{*}{\cellcolor[HTML]{F2F2F2}Did not mention}                                       \\
\cite{20193407357198}                                                   & Yes                                                                 & 40 seconds                                                                      & \begin{tabular}[c]{@{}l@{}}Dataset 1: \\ 81 depressed, 89   healthy\end{tabular}                                                                & 18 to 55 years                                                                                                                                         & MINI, LES, HAMD+CTQ                                                                                                                                        & \begin{tabular}[c]{@{}l@{}}Dataset 1 and 2: \\ Resting-state (eyes closed)\end{tabular}         \\
                                                           &                                                                     &                                                                                 &                                                                                                                                                 &                                                                                                                                                        &                                                                                                                                                            &                                                                                                 \\
\multirow{-2}{*}{}                                         & \multirow{-2}{*}{}                                                  & \multirow{-2}{*}{}                                                              & \multirow{-2}{*}{\begin{tabular}[c]{@{}l@{}}Dataset 2:\\ 160 depressed, 116 healthy\end{tabular}}                                               & \multirow{-2}{*}{}                                                                                                                                     & \multirow{-2}{*}{}                                                                                                                                         & \multirow{-2}{*}{\begin{tabular}[c]{@{}l@{}}Dataset 3 and 4: \\ auditory stimulus\end{tabular}} \\
\rowcolor[HTML]{F2F2F2} 
\cite{18951307}                                                   & Yes                                                                 & 4 minutes                                                                       & 60 (30F, 30M) depressed                                                                                                                         & 32.4±10.5                                                                                                                                            & \begin{tabular}[c]{@{}l@{}}Based on DSM-IV by a \\ psychiatrist and BDI-II\end{tabular}                                                                    & Resting-state (eyes closed)                                                                     \\
                                                           &                                                                     &                                                                                 & 30 depressed                                                                                                                                    &                                                                                                                                                        &                                                                                                                                                            &                                                                                                 \\
\multirow{-2}{*}{\cite{Sandheep2019}}                                 & \multirow{-2}{*}{Yes}                                               & \multirow{-2}{*}{5 minutes}                                                     & 30 healthy                                                                                                                                      & \multirow{-2}{*}{20 to 50 years}                                                                                                                       & \multirow{-2}{*}{Did not mention}                                                                                                                          & \multirow{-2}{*}{Eyes closed and eyes open}                                                     \\
\rowcolor[HTML]{F2F2F2} 
\cellcolor[HTML]{F2F2F2}                                   & \cellcolor[HTML]{F2F2F2}                                            & \cellcolor[HTML]{F2F2F2}                                                        & 33 (18F, 15M) depressed                                                                                                                         & Depressed: 40.33±12.861                                                                                                                              & \cellcolor[HTML]{F2F2F2}                                                                                                                                   & \cellcolor[HTML]{F2F2F2}                                                                        \\
\rowcolor[HTML]{F2F2F2} 
\multirow{-2}{*}{\cellcolor[HTML]{F2F2F2}\cite{Mumtaz2019}}         & \multirow{-2}{*}{\cellcolor[HTML]{F2F2F2}Yes}                       & \multirow{-2}{*}{\cellcolor[HTML]{F2F2F2}10 minutes}                            & 30 (9F, 21M) healthy                                                                                                                            & Healthy: 38.227±15.64                                                                                                                                & \multirow{-2}{*}{\cellcolor[HTML]{F2F2F2}DSM-IV, BDI-II and HADS}                                                                                          & \multirow{-2}{*}{\cellcolor[HTML]{F2F2F2}Eyes closed and eyes open}                             \\
\cite{20194907787800}                                                   & Yes                                                                 & 5 minutes                                                                       & 30 depressed                                                                                                                                    & 20 to 50 years                                                                                                                                         & Did not mention                                                                                                                                            & Eyes closed and eyes open                                                                       \\
\rowcolor[HTML]{F2F2F2} 
\cellcolor[HTML]{F2F2F2}                                   & \cellcolor[HTML]{F2F2F2}                                            & \cellcolor[HTML]{F2F2F2}                                                        & 10 depressed                                                                                                                                    & \cellcolor[HTML]{F2F2F2}                                                                                                                               & \cellcolor[HTML]{F2F2F2}                                                                                                                                   & \cellcolor[HTML]{F2F2F2}                                                                        \\
\rowcolor[HTML]{F2F2F2} 
\multirow{-2}{*}{\cellcolor[HTML]{F2F2F2}\cite{19258582}}         & \multirow{-2}{*}{\cellcolor[HTML]{F2F2F2}Yes}                       & \multirow{-2}{*}{\cellcolor[HTML]{F2F2F2}4 minutes}                             & 12 healthy                                                                                                                                      & \multirow{-2}{*}{\cellcolor[HTML]{F2F2F2}Did not mention}                                                                                              & \multirow{-2}{*}{\cellcolor[HTML]{F2F2F2}Did not mention}                                                                                                  & \multirow{-2}{*}{\cellcolor[HTML]{F2F2F2}Eyes closed and eyes open}                             \\
                                                           &                                                                     &                                                                                 & 27 (11F, 16M) depressed                                                                                                                         & Depressed: 31.67±10.94                                                                                                                               &                                                                                                                                                            &                                                                                                 \\
\multirow{-2}{*}{\cite{Peng2019}}                                & \multirow{-2}{*}{Yes}                                               & \multirow{-2}{*}{5 minutes}                                                     & 28 (9F, 19M) healthy                                                                                                                            & Healthy: 31.82±8.76                                                                                                                                  & \multirow{-2}{*}{\begin{tabular}[c]{@{}l@{}}PHQ-9 and \\ GAD-7 by clinical psychiatrists\end{tabular}}                                                     & \multirow{-2}{*}{Resting-state (eyes closed)}                                                   \\
\rowcolor[HTML]{F2F2F2} 
\cellcolor[HTML]{F2F2F2}                                   & \cellcolor[HTML]{F2F2F2}                                            & \cellcolor[HTML]{F2F2F2}                                                        & 24 (6F, 18M) depressed                                                                                                                          & Depressed: 20.96±1.95                                                                                                                                & \cellcolor[HTML]{F2F2F2}                                                                                                                                   & \cellcolor[HTML]{F2F2F2}                                                                        \\
\rowcolor[HTML]{F2F2F2} 
\multirow{-2}{*}{\cellcolor[HTML]{F2F2F2}\cite{18927189}}         & \multirow{-2}{*}{\cellcolor[HTML]{F2F2F2}Yes}                       & \multirow{-2}{*}{\cellcolor[HTML]{F2F2F2}Did not mention}                       & 24 (9F, 15M) healthy                                                                                                                            & Healthy: 20±2.02                                                                                                                                     & \multirow{-2}{*}{\cellcolor[HTML]{F2F2F2}\begin{tabular}[c]{@{}l@{}}Psychological screening and \\ interview by specialist and BDI-II\end{tabular}}        & \multirow{-2}{*}{\cellcolor[HTML]{F2F2F2}Eyes open with visual stimulation}                     \\
\cite{19203658}                                                   & Yes                                                                 & Experiment 1: 8 minutes                                                         & \begin{tabular}[c]{@{}l@{}}Experiment 1: \\ 23 (12F, 11M)  depressed, \\ 12 (6F, 6M) healthy\end{tabular}                                       & \begin{tabular}[c]{@{}l@{}}Experiment 1 \\ (healthy: 26.4±9.8, \\ depressed: 29.3±9.6),\end{tabular}                                               & DSM-IV, HAM-D17, MINI                                                                                                                                      & Resting-state (eyes closed)                                                                     \\
                                                           &                                                                     &                                                                                 &                                                                                                                                                 &                                                                                                                                                        &                                                                                                                                                            &                                                                                                 \\
\multirow{-2}{*}{}                                         & \multirow{-2}{*}{}                                                  & \multirow{-2}{*}{Experiment 2: 8 minutes}                                       & \multirow{-2}{*}{\begin{tabular}[c]{@{}l@{}}Experiment 2:\\ 15 depressed, 15 healthy\end{tabular}}                                              & \multirow{-2}{*}{\begin{tabular}[c]{@{}l@{}}Experiment 2 \\ (healthy: 33.6±11.5, \\ depressed: 34±9.2)\end{tabular}}                               & \multirow{-2}{*}{}                                                                                                                                         & \multirow{-2}{*}{}                                                                              \\
\rowcolor[HTML]{F2F2F2} 
\cellcolor[HTML]{F2F2F2}                                   & \cellcolor[HTML]{F2F2F2}                                            & \cellcolor[HTML]{F2F2F2}                                                        & 34 (17F, 17M) depressed                                                                                                                         & Depressed: 40.3±12.9                                                                                                                                 & \cellcolor[HTML]{F2F2F2}                                                                                                                                   & \cellcolor[HTML]{F2F2F2}                                                                        \\
\rowcolor[HTML]{F2F2F2} 
\multirow{-2}{*}{\cellcolor[HTML]{F2F2F2}\cite{20183405723339}}         & \multirow{-2}{*}{\cellcolor[HTML]{F2F2F2}No}                        & \multirow{-2}{*}{\cellcolor[HTML]{F2F2F2}5 minutes}                             & 30 (9F, 21M) healthy                                                                                                                            & Healthy: 38.3±15.6                                                                                                                                   & \multirow{-2}{*}{\cellcolor[HTML]{F2F2F2}\begin{tabular}[c]{@{}l@{}}DSM-IV criteria \\ (American psychiatric association 1994)\end{tabular}}               & \multirow{-2}{*}{\cellcolor[HTML]{F2F2F2}Resting-state (eyes closed)}                           \\
                                                           &                                                                     &                                                                                 & 16 (8F, 8M) depressed                                                                                                                           & Depressed: 43.44±13.27                                                                                                                               &                                                                                                                                                            &                                                                                                 \\
\multirow{-2}{*}{\cite{20190906547554}}                                & \multirow{-2}{*}{Yes}                                               & \multirow{-2}{*}{4 minutes}                                                     & 16 (8F, 8M) healthy                                                                                                                             & Healthy: 43.19±13.03                                                                                                                                 & \multirow{-2}{*}{DSM, HAMD}                                                                                                                                & \multirow{-2}{*}{Resting-state (eyes closed)}

\end{tabular}}
\end{sidewaystable*}

\begin{sidewaystable*}[htp]
\caption{\label{tab:Table3-2} Continued - summary of the most recent literature categorized based on dataset, experiment duration, number of subjects and their gender, age, selection criteria and condition of the experiment}
\resizebox{1\textwidth}{!}{\begin{tabular}{lllllllll}

\multicolumn{1}{c}{}                                       & \multicolumn{1}{c}{}                                                & \multicolumn{1}{c}{}                                                            & \multicolumn{1}{c}{}                                                                                                                            &                                                                                                                                                        & \multicolumn{1}{c}{}                                                                                                                                       & \multicolumn{1}{c}{}                                                                            \\
\multicolumn{1}{c}{\multirow{-2}{*}{\textit{\textbf{\#}}}} & \multicolumn{1}{c}{\multirow{-2}{*}{\textit{\textbf{Own dataset}}}} & \multicolumn{1}{c}{\multirow{-2}{*}{\textit{\textbf{Duration (for analysis)}}}} & \multicolumn{1}{c}{\multirow{-2}{*}{\textit{\textbf{Subjects/Gender}}}}                                                                         & \multirow{-2}{*}{\textit{\textbf{Age (Mean ± SD)}}}                                                                                                   & \multicolumn{1}{c}{\multirow{-2}{*}{\textit{\textbf{Participant selection and rating scale}}}}                                                             & \multicolumn{1}{c}{\multirow{-2}{*}{\textit{\textbf{Experiment condition}}}}                    \\

\rowcolor[HTML]{F2F2F2} 
\cite{18983501}            & Yes                           & 5 minutes                                 & 15 depressed                                                                                          & 20 to 50 years              & Diagnosis by psychiatrists                                                                                                                  & Eyes closed and eyes open                          \\
\rowcolor[HTML]{F2F2F2} 
                     &                               &                                           & 15 healthy                                                                                            &                             &                                                                                                                                             &                                                    \\
\cite{20185106271469}             & Yes                           & 5 minutes                                 & 17 (6F, 11M) depressed                                                                                & Depressed: 33.35±12.36    & MINI and PHQ-9 with the help of Psychiatrists                                                                                               & Resting-state                                      \\
                     &                               &                                           & 17 (4F, 13M) healthy                                                                                  & Healthy: 30.29±9.68       &                                                                                                                                             &                                                    \\
\rowcolor[HTML]{F2F2F2} 
\cite{18853691}             & Yes                           & 5 minutes                                 & 9 depressed                                                                                           & 18 to 60 years              & DSM-IV, HAM-D17, MINI                                                                                                                       & Resting-state (eyes closed)                        \\
\cite{Cai2018ACASE}            & Yes                           & 6 minutes                                 & 86 depressed                                                                                          & 18 to 55 years              & \begin{tabular}[c]{@{}l@{}}PHQ-9, LES, PSQI, GDA-7, SCSQ, SSRS, \\ EPQ-R short scale, CAQ, \\ Social information questionnaire\end{tabular} & Resting and audio stimulation state                \\
                     &                               &                                           & 92 healthy                                                                                            &                             &                                                                                                                                             &                                                    \\
\rowcolor[HTML]{F2F2F2} 
\cite{Cai2018ADetection}            & Yes                           & 6 minutes                                 & 92 depressed                                                                                          & Did not mention             & \begin{tabular}[c]{@{}l@{}}Interview by a psychiatrist, PHQ-9,\\ LES, PSQI, GAD-7\end{tabular}                                              & Resting and audio stimulation state                \\
\rowcolor[HTML]{F2F2F2} 
                     &                               &                                           & 121 healthy                                                                                           &                             &                                                                                                                                             &                                                    \\
\cite{Acharya2018AutomatedNetwork}             & No                            & 5 minutes                                 & 15 depressed                                                                                          & 20 to 50 years              & Questionnaire and physical examination                                                                                                      & Eyes closed and eyes open                          \\
                     &                               &                                           & 15 healthy                                                                                            &                             &                                                                                                                                             &                                                    \\
\rowcolor[HTML]{F2F2F2} 
\cite{20184806131498}             & Yes                           & Dataset1: 72 seconds                      & \begin{tabular}[c]{@{}l@{}}Dataset 1: \\ 34 (17F, 17M) depressed, \\ 30 (9F,21M) healthy\end{tabular} & Dataset1:                   & Screened for possible mental or physical illness                                                                                            & Did not mention                                    \\
\rowcolor[HTML]{F2F2F2} 
                     &                               & Dataset 2: 48 seconds                     & Dataset 2: 17 depressed                                                                               & depressed: 40.3±12.9      &                                                                                                                                             &                                                    \\
\rowcolor[HTML]{F2F2F2} 
                     &                               &                                           &                                                                                                       & healthy: 38.227±15.64     &                                                                                                                                             &                                                    \\
\cite{Bachmann2018}             & Yes                           & 5 minutes eyes-closed                     & 13 (8F, 5M) depressed                                                                                 & 38.7                        & ICD-10, EST-Q                                                                                                                               & Eyes closed and eyes open                          \\
                     &                               &                                           & 13 (8F, 5M) healthy                                                                                   & 15.8                        &                                                                                                                                             &                                                    \\
\rowcolor[HTML]{F2F2F2} 
{[}61{]}             & Yes                           & 10 minutes                                & 34 (18F, 16M) depressed                                                                               & Depressed: 40.33±12.861   & DSM-IV, BDI-II, HADS                                                                                                                        & Eyes closed and eyes open                          \\
\rowcolor[HTML]{F2F2F2} 
                     &                               &                                           & 30 (9F, 21M) healthy                                                                                  & Healthy: 38.227±15.64     &                                                                                                                                             &                                                    \\
\cite{20173504090707}           & Yes                           & 3 minutes                                 & 3 depressed                                                                                           & Did not mention             & Did not mention                                                                                                                             & Did not mention                                    \\
                     &                               &                                           & 3 healthy                                                                                             &                             &                                                                                                                                             &                                                    \\
\rowcolor[HTML]{F2F2F2} 
\cite{Wan2017AQA}             & Yes                           & 5 minutes                                 & 6 depressed                                                                                           & Did not mention             & Did not mention                                                                                                                             & Resting-state (eyes closed)                        \\
\cite{Shen2017ACriterion}             & Yes                           & 40 seconds                                & 81 depressed                                                                                          & 18 to 55 years              & PHQ-9, CTQ, MINI, LES                                                                                                                       & Resting-state (eyes closed)                        \\
                     &                               &                                           & 89 healthy                                                                                            &                             &                                                                                                                                             &                                                    \\
\rowcolor[HTML]{F2F2F2} 
\cite{Zhao2017}            & Yes                           & 72 seconds audio stimulation              & 81 depressed                                                                                          & Did not mention             & PHQ-9, MINI                                                                                                                                 & Resting and audio stimulation state                \\
\rowcolor[HTML]{F2F2F2} 
                     &                               &                                           & 89 healthy                                                                                            &                             &                                                                                                                                             &                                                    \\
\cite{Wan2017}            & Yes                           & Not mentioned                             & 5 depressed                                                                                           & Did not mention             & \begin{tabular}[c]{@{}l@{}}POMS-BCN, HAMD-17, YMRS, PHQ-9, \\ GAD-7, LES, YMRS, MINI\end{tabular}                                           & Resting-state                                      \\
\rowcolor[HTML]{F2F2F2} 
\cite{Ritu2017}            & Yes                           & 14 minutes                                & 10 depressed,                                                                                         & 20.275±0.798              & DASS-21                                                                                                                                     & Resting-state, audio, and visual stimulation       \\
\rowcolor[HTML]{F2F2F2} 
                     &                               &                                           & 10 healthy                                                                                            &                             &                                                                                                                                             &                                                    \\
\cite{17991812}            & Yes                           & 4 minutes                                 & 75 depressed                                                                                          & Did not mention             & DSM-IV, BDI                                                                                                                                 & Did not mention                                    \\
\rowcolor[HTML]{F2F2F2} 
\cite{Orgo2017}            & Yes                           & 6 minutes                                 & 37 (21F, 16M) depressed                                                                               & Depressed: 33.9±14.3      & Diagnosis by physicians                                                                                                                     & Resting-state (eyes closed)                        \\
\rowcolor[HTML]{F2F2F2} 
                     &                               &                                           & 37 (21F, 16M) healthy                                                                                 & Healthy: 33.9±14.6        &                                                                                                                                             &                                                    \\
\cite{20183605764415}             & Yes                           & 4 minutes                                 & 16 (8F, 8M) depressed                                                                                 & 40                          & Did not mention                                                                                                                             & Resting-state (Eyes closed)                        \\
                     &                               &                                           & 16 (8F, 8M) healthy                                                                                   & 28.2                        &                                                                                                                                             &                                                    \\
\rowcolor[HTML]{F2F2F2} 
\cite{17051427}              & Yes                           & 5 minutes                                 & 23 (10F, 13M) depressed                                                                               & Depressed: 33.17±19.83    & MINI, PHQ-9                                                                                                                                 & Resting-state (Eyes closed)                        \\
\rowcolor[HTML]{F2F2F2} 
                     &                               &                                           & 14 (7F, 7M) healthy                                                                                   & Healthy: 31.29±21.71      &                                                                                                                                             &                                                    \\
\cite{Cai2017NO2}             & Yes                           & 84 seconds                                & 90 (50F, 40M) depressed                                                                               & Did not mention             & MINI, DSM-IV, PHQ-9                                                                                                                         & Resting-state (eyes closed) with audio stimulation \\
                     &                               &                                           & 68 (34F, 34M) healthy                                                                                 &                             &                                                                                                                                             &                                                    \\
\rowcolor[HTML]{F2F2F2} 
\cite{Liao2017}            & Yes                           & 5 minutes and 24 seconds                  & 12 (8F, 4M) depressed                                                                                 & Depressed: 57±7.71        & DSM-IV-TR, MINI                                                                                                                             & Resting-state (eyes open)                          \\
\rowcolor[HTML]{F2F2F2} 
                     &                               &                                           & 12 (8F, 4M) healthy                                                                                   & Healthy: 57±7.71          &                                                                                                                                             &

\end{tabular}}
\end{sidewaystable*}

\begin{sidewaystable*}[htp]
\caption{\label{tab:Table3-3} Continued - summary of the most recent literature categorized based on dataset, experiment duration, number of subjects and their gender, age, selection criteria and condition of the experiment}
\resizebox{1\textwidth}{!}{\begin{tabular}{lllllllll}

\multicolumn{1}{c}{}                                       & \multicolumn{1}{c}{}                                                & \multicolumn{1}{c}{}                                                            & \multicolumn{1}{c}{}                                                                                                                            &                                                                                                                                                        & \multicolumn{1}{c}{}                                                                                                                                       & \multicolumn{1}{c}{}                                                                            \\
\multicolumn{1}{c}{\multirow{-2}{*}{\textit{\textbf{\#}}}} & \multicolumn{1}{c}{\multirow{-2}{*}{\textit{\textbf{Own dataset}}}} & \multicolumn{1}{c}{\multirow{-2}{*}{\textit{\textbf{Duration (for analysis)}}}} & \multicolumn{1}{c}{\multirow{-2}{*}{\textit{\textbf{Subjects/Gender}}}}                                                                         & \multirow{-2}{*}{\textit{\textbf{Age (Mean ± SD)}}}                                                                                                   & \multicolumn{1}{c}{\multirow{-2}{*}{\textit{\textbf{Participant selection and rating scale}}}}                                                             & \multicolumn{1}{c}{\multirow{-2}{*}{\textit{\textbf{Experiment condition}}}}                    \\

\rowcolor[HTML]{F2F2F2} 
\cite{Mumtaz2017}             & Yes                           & 2 minutes                                 & 33 depressed                        & Depressed: 40.33±12.861       & DSM-IV, BDI-II, HADS                                                                                                                                         & Eyes closed and eyes open                                                                                   \\
\rowcolor[HTML]{F2F2F2} 
                     &                               &                                           & 30 healthy                          & Healthy: 38.227 ± 15.64         &                                                                                                                                                              &                                                                                                             \\
\cite{Puthankattil2017}             & Yes                           & 5 minutes                                 & 30 (16F, 14M) depressed             & Female: 33 years                & Did not mention                                                                                                                                              & Resting-state (eyes closed and open)                                                                        \\
                     &                               &                                           & 30 (16F, 14M) healthy               & Male: 35 years                  &                                                                                                                                                              &                                                                                                             \\
\rowcolor[HTML]{F2F2F2} 
\cite{Cai2016PervasiveCollector}             & Yes                           & 6 minutes                                 & 86 depressed                        & 18 to 55 years                  & \begin{tabular}[c]{@{}l@{}}PHQ-9, LES, PSQI, GAD-7, MINI, \\ CTQ, SCSQ, SSRS, EPQ-R short scale, \\ Social information processing questionnaire\end{tabular} & Resting-state with audio stimulation                                                                        \\
\rowcolor[HTML]{F2F2F2} 
                     &                               &                                           & 92 healthy                          &                                 &                                                                                                                                                              &                                                                                                             \\
\cite{Orgo2017-NO2}             & Yes                           & 6 minutes                                 & 37 (21F, 16M) depressed             & Depressed: 33.9±14.3          & Diagnosis by physicians                                                                                                                                      & Resting-state (eyes closed)                                                                                 \\
                     &                               &                                           & 37 (21F, 16M) healthy               & Healthy: 33.9±14.6            &                                                                                                                                                              &                                                                                                             \\
\rowcolor[HTML]{F2F2F2} 
\cite{Bachmann2017}             & Yes                           & 5 minutes                                 & 17 (17F) depressed                  & 39 ± 12                         & ICD-10, HAM-D17                                                                                                                                              & Resting-state (eyes closed and ears blocked)                                                                \\
\rowcolor[HTML]{F2F2F2} 
                     &                               &                                           & 17 (17F) healthy                    &                                 &                                                                                                                                                              &                                                                                                             \\
\cite{Li2016}             & Yes                           & 5 minutes                                 & 10 depressed                        & Mildly depressed: 20.60±1.838 & BDI-II                                                                                                                                                       & Eyes open with visual stimulation                                                                           \\
                     &                               &                                           & 10 (2F, 8M) Healthy                 & Healthy: 20.20±2.044          &                                                                                                                                                              &                                                                                                             \\
\rowcolor[HTML]{F2F2F2} 
{}\cite{Puk2016}{}              & Yes                           & Did not mention                           & 15 (11F, 4M) depressed              & Depressed: 20.3±3.21          & CES-D                                                                                                                                                        & Eyes open                                                                                                   \\
\rowcolor[HTML]{F2F2F2} 
                     &                               &                                           & 12 (8F, 4M) healthy                 & Healthy: 20.5±2.66            &                                                                                                                                                              &                                                                                                             \\
\cite{Mohan2016}             & Yes                           & 32.25 seconds                             & 63 depressed                        & 22.7±1.47                     & PHQ-9, DASS-21                                                                                                                                               & Did not mention                                                                                             \\
                     &                               &                                           & 53 healthy                          &                                 &                                                                                                                                                              &                                                                                                             \\
\rowcolor[HTML]{F2F2F2} 
\cite{XiaoweiLi2016EEG-basedClassifiers}             & Yes                           & 5 minutes                                 & 10 (4F, 6M) depressed               & Depressed: 20.60 ± 1.74         & Questionnaire survey, BDI-II                                                                                                                                 & Eyes open                                                                                                   \\
\rowcolor[HTML]{F2F2F2} 
                     &                               &                                           & 10 (2F, 8M) healthy                 & Healthy: 20.3±1.85            &                                                                                                                                                              &                                                                                                             \\
\cite{Li2016-NO2}           & Yes                           & 6 minutes                                 & 14 depressed                        & Depressed: 20.78±1.75         & BDI-II, OASIS, K10                                                                                                                                           & Eyes open                                                                                                   \\
                     &                               &                                           & 14 healthy                          & Healthy: 19.22±1.47           &                                                                                                                                                              &                                                                                                             \\
\rowcolor[HTML]{F2F2F2} 
\cite{Spyrou2016}             & Yes                           & 20 seconds                                & 34 depressed                        & Depressed: 69.818               & Interview, MoCA, MMSE, GDS,                                                                                                                                  & Resting-state (eyes closed)                                                                                 \\
\rowcolor[HTML]{F2F2F2} 
                     &                               &                                           & 32 healthy                          & Healthy: 70.333                 &                                                                                                                                                              &                                                                                                             \\
\cite{Erguzel2016}             & Yes                           & 2 minutes                                 & 31 bipolar depressed                & Did not mention                 & DSM-IV, SCID-I, HAMD-17, YMRS                                                                                                                                & Resting-state (eyes closed)                                                                                 \\
                     &                               &                                           & 58 unipolar depressed               &                                 &                                                                                                                                                              &                                                                                                             \\
\rowcolor[HTML]{F2F2F2} 
\cite{20160201780626}            & Yes                           & 3 minutes                                 & 16 (8F, 8M) depressed               & Depressed: 30.92±3.56         & DSM-IV-TR, HAMD                                                                                                                                              & Resting-state (eyes closed)                                                                                 \\
\rowcolor[HTML]{F2F2F2} 
                     &                               &                                           & 15 (7F, 8M) healthy                 & Healthy: 29.42±4.02           &                                                                                                                                                              &                                                                                                             \\
\cite{15605052}             & Yes                           & 15 Minutes                                & 15 (7F, 8M) depressed               & Depressed: 30.92±3.65         & DSM-IV, HAM-D                                                                                                                                                & Resting-state with audio stimulation                                                                        \\
                     &                               &                                           & 15 (7F, 8M) healthy                 & Healthy: 29.42±4.02           &                                                                                                                                                              &                                                                                                             \\
\rowcolor[HTML]{F2F2F2} 
\cite{Kalev2015}             & Yes                           & 34 minutes                                & 11 depressed                        & 42.5±14.8                     & ICD-10                                                                                                                                                       & Resting-state (eyes closed and open)                                                                        \\
\rowcolor[HTML]{F2F2F2} 
                     &                               &                                           & 11 healthy                          &                                 &                                                                                                                                                              &                                                                                                             \\
\rowcolor[HTML]{F2F2F2} 
                     &                               &                                           & (12F, 10M)                          &                                 &                                                                                                                                                              &                                                                                                             \\
\cite{20161302158441}             & Yes                           & 4 minutes                                 & 4 depressed                         & 23 to 25 years (Mean: 23.38)    & PHQ-9, DASS-21                                                                                                                                               & Resting-state (eyes closed and open)                                                                        \\
                     &                               &                                           & 4 healthy                           &                                 &                                                                                                                                                              &                                                                                                             \\
                     &                               &                                           & (3F, 5M)                            &                                 &                                                                                                                                                              &                                                                                                             \\
\rowcolor[HTML]{F2F2F2} 
\cite{15798207}             & Yes                           & 3 minutes                                 & 9 (4F, 5M) depressed                & Did not mention                 & BDI-II                                                                                                                                                       & Eyes open with viewing task                                                                                 \\
\rowcolor[HTML]{F2F2F2} 
                     &                               &                                           & 25 healthy                          &                                 &                                                                                                                                                              &                                                                                                             \\
\cite{Li2015}             & Yes                           & 6 minutes                                 & 9 depressed                         & Depressed: 20.78±1.85         & OASIS, K10                                                                                                                                                   & Eyes open with viewing task                                                                                 \\
                     &                               &                                           & 9 healthy                           & Healthy: 20.00±2.78           &                                                                                                                                                              &                                                                                                             \\
                     &                               &                                           & (3F, 15M)                           &                                 &                                                                                                                                                              &                                                                                                             \\
\rowcolor[HTML]{F2F2F2} 
\cite{15670461}            & Yes                           & 5 minutes                                 & 13 depressed                        & 16 to 66 years                  & DSM-IV                                                                                                                                                       & Resting-state                                                                                               \\
\rowcolor[HTML]{F2F2F2} 
                     &                               &                                           & 12 healthy                          &                                 &                                                                                                                                                              &                                                                                                             \\
\rowcolor[HTML]{F2F2F2} 
                     &                               &                                           & (13F, 12M)                          &                                 &                                                                                                                                                              &                                                                                                             \\
                     
\end{tabular}}
\end{sidewaystable*}

\begin{sidewaystable*}[htp]
\caption{\label{tab:Table3-4} Continued - summary of the most recent literature categorized based on dataset, experiment duration, number of subjects and their gender, age, selection criteria and condition of the experiment}
\resizebox{1\textwidth}{!}{\begin{tabular}{lllllllll}

\multicolumn{1}{c}{}                                       & \multicolumn{1}{c}{}                                                & \multicolumn{1}{c}{}                                                            & \multicolumn{1}{c}{}                                                                                                                            &                                                                                                                                                        & \multicolumn{1}{c}{}                                                                                                                                       & \multicolumn{1}{c}{}                                                                            \\
\multicolumn{1}{c}{\multirow{-2}{*}{\textit{\textbf{\#}}}} & \multicolumn{1}{c}{\multirow{-2}{*}{\textit{\textbf{Own dataset}}}} & \multicolumn{1}{c}{\multirow{-2}{*}{\textit{\textbf{Duration (for analysis)}}}} & \multicolumn{1}{c}{\multirow{-2}{*}{\textit{\textbf{Subjects/Gender}}}}                                                                         & \multirow{-2}{*}{\textit{\textbf{Age (Mean ± SD)}}}                                                                                                   & \multicolumn{1}{c}{\multirow{-2}{*}{\textit{\textbf{Participant selection and rating scale}}}}                                                             & \multicolumn{1}{c}{\multirow{-2}{*}{\textit{\textbf{Experiment condition}}}}                    \\

\cite{Mumtaz2015-NO2}            & Yes                           & 10 minutes                                & 33 (17F, 16M) depressed             & Depressed: 40.33±12.861        & DSM-IV, BDI-II                                                                                                                    & Resting-state (eyes closed and open)                                                                        \\
                     &                               &                                           & 30 (9F, 21M) healthy                & Healthy: 38.27±15.64,          &                                                                                                                                   &                                                                                                             \\
\rowcolor[HTML]{F2F2F2} 
{}\cite{Mumtaz2015ADisorder}{}              & Yes                           & 10 minutes                                & 33 depressed                        & Depressed: 40.33±12.86       & DSM-IV                                                                                                                            & Resting-state (eyes closed and open)                                                                        \\
\rowcolor[HTML]{F2F2F2} 
                     &                               &                                           & 19 healthy                          & Healthy: 39.8±15.6           &                                                                                                                                   &                                                                                                             \\
{}\cite{TekinErguzel2015}{}              & Yes                           & 3 minutes                                 & 46 (29F, 17M) bipolar disorder      & Did not mention                & DSM-IV, SCID-I, HAM-D17, YMRS                                                                                                     & Resting-state (eyes closed)                                                                                 \\
                     &                               &                                           & 55 (32F, 23M) MDD                   &                                &                                                                                                                                   &                                                                                                             \\
\rowcolor[HTML]{F2F2F2} 
\cite{Sun2015}            & Yes                           & 90 Seconds                                & 11 depressed                        & Average: 30 years              & \begin{tabular}[c]{@{}l@{}}EPQ, PSQI, LES, GAD-7, PHQ-9, CAQ, \\ SCSQ, MINI, \\ Amount of Social Support Composition\end{tabular} & Resting-state (eyes closed)                                                                                 \\
\rowcolor[HTML]{F2F2F2} 
                     &                               &                                           & 11 healthy                          &                                &                                                                                                                                   &                                                                                                             \\
\cite{Mumtaz2016}            & Yes                           & 20 minutes                                & 29 (14F, 15M) depressed             & Depressed: 40.33±12.861      & DSM-IV                                                                                                                            & \begin{tabular}[c]{@{}l@{}}Resting-state (eyes closed and eyes open) and \\ visual stimulation\end{tabular} \\
                     &                               &                                           & 15 (9F, 6M) healthy                 & Healthy: 38.277±15.64        &                                                                                                                                   &                                                                                                             \\
\rowcolor[HTML]{F2F2F2} 
{}\cite{Bachmann2015}{}              & Yes                           & 5 minutes                                 & 17 (17F) depressed                  & 39±12                        & ICD-10, HAM-D                                                                                                                     & Resting-state (eyes closed and ears blocked)                                                                \\
\rowcolor[HTML]{F2F2F2} 
                     &                               &                                           & 17 (17F) healthy                    &                                &                                                                                                                                   &                                                                                                             \\
\cite{Katyal2014}             & No                            & Did not mention                           & 10 depressed                        & Did not mention                & Did not mention                                                                                                                   & Did not mention                                                                                             \\
                     &                               &                                           & 10 healthy                          &                                &                                                                                                                                   &                                                                                                             \\
\rowcolor[HTML]{F2F2F2} 
\cite{Faust2014}            & Yes                           & 5 minutes                                 & 30 (16F, 14M) depressed             & 20 to 50 years                 & Did not mention                                                                                                                   & Resting-state                                                                                               \\
\rowcolor[HTML]{F2F2F2} 
                     &                               &                                           & 30 (16F, 14M) healthy               &                                &                                                                                                                                   &                                                                                                             \\
\cite{Zhao2013}             & Yes                           & Did not mention                           & 22 (10F, 12M) depressed             & Depressed: 31.7±9.5        & Did not mention                                                                                                                   & Eyes open with facial expression stimulation                                                                \\
                     &                               &                                           & 22 (13F, 9M) normal                 & Healthy: 33.9±8.7            &                                                                                                                                   &                                                                                                             \\
\rowcolor[HTML]{F2F2F2} 
\cite{13386381}             & Yes                           & 1 minute every day for a month            & 13 (13F) depressed                  & 30 to 40 years                 & BDI-II                                                                                                                            & Resting-state (eyes closed)                                                                                 \\
\rowcolor[HTML]{F2F2F2} 
                     &                               &                                           & 12 (12F) healthy                    &                                &                                                                                                                                   &                                                                                                             \\
{}\cite{Liao2013}{}             & Yes                           & 8 minutes                                 & 7 (2F, 5M) unmedicated depressed    & Unmedicated depressed:         & DSM-IV, BDI, CGI-S, HAMD, QIDS-SR, T-AI                                                                                           & Resting-state (eyes closed)                                                                                 \\
                     &                               &                                           & 5 (2F, 3M) medicated depressed      & 30.43±6.63                  &                                                                                                                                   &                                                                                                             \\
                     &                               &                                           & 10 (5F, 5M) healthy                 & Medicated depressed:           &                                                                                                                                   &                                                                                                             \\
                     &                               &                                           &                                     & 46.25±8.88                  &                                                                                                                                   &                                                                                                             \\
                     &                               &                                           &                                     & Healthy:                       &                                                                                                                                   &                                                                                                             \\
                     &                               &                                           &                                     & 23.32±2.55                  &                                                                                                                                   &                                                                                                             \\
\rowcolor[HTML]{F2F2F2} 
\cite{Suhhova2013}            & Yes                           & 10 minutes                                & Group 1 and 2: 18 (18F) depressed   & Group 1 and 2: 35±11         & ICD-10, HAM-D                                                                                                                     & Resting-state (eyes closed)                                                                                 \\
\rowcolor[HTML]{F2F2F2} 
                     &                               &                                           & 18 (18F) healthy                    & Group 3: 22±1                &                                                                                                                                   &                                                                                                             \\
\rowcolor[HTML]{F2F2F2} 
                     &                               &                                           & Group 3 and 4: 15 healthy           & Group 4: 27±4                &                                                                                                                                   &                                                                                                             \\
{}\cite{Hosseinifard2013}{}             & Yes                           & 5 minutes                                 & 45 (23F, 22M) unmedicated depressed & Depressed: 33.5±10.7         & DSM-IV, BDI                                                                                                                       & Resting-state (eyes closed)                                                                                 \\
                     &                               &                                           & 45 (25F, 20M) healthy               & Healthy: 33.7±10.2           &                                                                                                                                   &                                                                                                             \\
\rowcolor[HTML]{F2F2F2} 
\cite{Frantzidis2012}             & Yes                           & 4 minutes                                 & 33 (33M) depressed                  & Depressed: 69.818              & GDS, MMSE, MoCA                                                                                                                   & Resting-state (eyes closed)                                                                                 \\
\rowcolor[HTML]{F2F2F2} 
                     &                               &                                           & 33 (33M) healthy                    & Healthy: 70.333                &                                                                                                                                   &                                                                                                             \\
\cite{Puthankattil2012ClassificationEntropy}             & Yes                           & 5 minutes                                 & 30 (16F, 14M) depressed             & 20 to 50 years                 & Did not mention                                                                                                                   & Resting-state                                                                                               \\
                     &                               &                                           & 30 (16F, 14M) healthy               &                                &                                                                                                                                   &                                                                                                             \\
\rowcolor[HTML]{F2F2F2} 
\cite{Ku2012}             & Yes                           & 5 + 21.6 minutes under T.O.V.A            & 6 (6F) depressed                    & Depressed:                     & Did not mention                                                                                                                   & Did not mention                                                                                             \\
\rowcolor[HTML]{F2F2F2} 
                     &                               &                                           & 8 (8F) healthy                      & 46.6±8.63                   &                                                                                                                                   &                                                                                                             \\
\rowcolor[HTML]{F2F2F2} 
                     &                               &                                           &                                     & Healthy:                       &                                                                                                                                   &                                                                                                             \\
\rowcolor[HTML]{F2F2F2} 
                     &                               &                                           &                                     & 23.37±2.78                  &                                                                                                                                   &                                                                                                             \\
{}\cite{12424821}{}             & Yes                           & 5 minutes                                 & 12 (5F, 7M) depressed               & Depressed: 37.2±11.8         & \begin{tabular}[c]{@{}l@{}}CCMD-3 (F32: depressive episode, \\ Chinese Psychiatric Association, 2001), ICD-10\end{tabular}        & Resting-state (eyes closed)                                                                                 \\
                     &                               &                                           & 12 (5F, 7M) healthy                 & Healthy: 37.5±11.7           &                                                                                                                                   &                                                                                                             \\
\rowcolor[HTML]{F2F2F2} 
\cite{20102413002260}              & Yes                           & 30 minutes                                & 18 (18F) depressed                  & Depressed: 36±10             & HAM-D, ICD-10                                                                                                                     & Resting-state (eyes closed)                                                                                 \\
\rowcolor[HTML]{F2F2F2} 
                     &                               &                                           & 18 (18F) healthy                    & Healthy: 35±10.5             &                                                                                                                                   &                                                                                                             \\
                                                                                    
\end{tabular}}
\end{sidewaystable*}

\begin{sidewaystable*}[htp]
\caption{\label{tab:Table3-5} Continued - summary of the most recent literature categorized based on dataset, experiment duration, number of subjects and their gender, age, selection criteria and condition of the experiment}
\resizebox{1\textwidth}{!}{\begin{tabular}{lllllllll}

\multicolumn{1}{c}{}                                       & \multicolumn{1}{c}{}                                                & \multicolumn{1}{c}{}                                                            & \multicolumn{1}{c}{}                                                                                                                            &                                                                                                                                                        & \multicolumn{1}{c}{}                                                                                                                                       & \multicolumn{1}{c}{}                                                                            \\
\multicolumn{1}{c}{\multirow{-2}{*}{\textit{\textbf{\#}}}} & \multicolumn{1}{c}{\multirow{-2}{*}{\textit{\textbf{Own dataset}}}} & \multicolumn{1}{c}{\multirow{-2}{*}{\textit{\textbf{Duration (for analysis)}}}} & \multicolumn{1}{c}{\multirow{-2}{*}{\textit{\textbf{Subjects/Gender}}}}                                                                         & \multirow{-2}{*}{\textit{\textbf{Age (Mean ± SD)}}}                                                                                                   & \multicolumn{1}{c}{\multirow{-2}{*}{\textit{\textbf{Participant selection and rating scale}}}}                                                             & \multicolumn{1}{c}{\multirow{-2}{*}{\textit{\textbf{Experiment condition}}}}                    \\

\rowcolor[HTML]{F2F2F2} 
\cite{Suhhova2009}             & Yes                           & 10 minutes                                & 18 (18F) depressed                & 39±10                        & ICD-10, HAM-D                                                                                      & Resting-state (eyes closed and ears blocked) \\
\rowcolor[HTML]{F2F2F2} 
                     &                               &                                           & 18 (18F) healthy                  &                                &                                                                                                    &                                              \\
\cite{Hinrikus2009}            & Yes                           & 30 Minutes                                & 18 (18F) depressed                & 35±11                        & ICD-10, HAM-D                                                                                      & Resting-state (eyes closed and ears blocked) \\
                     &                               &                                           & 18 (18F) healthy                  &                                &                                                                                                    &                                              \\
\rowcolor[HTML]{F2F2F2} 
\cite{Lee2007}            & Yes                           & 5 minutes                                 & 11 (9F, 2M) unipolar depressed    & 44±11                        & BDI, HAM-D17, DSM-IV                                                                               & Resting-state (eyes closed)                  \\
\rowcolor[HTML]{F2F2F2} 
                     &                               &                                           & 11 (9F, 2M) healthy               &                                &                                                                                                    &                                              \\
\cite{Li2007}            & Yes                           & 5 minutes                                 & 26 (14F, 12M) depressed           & Depressed: 37.1                & \begin{tabular}[c]{@{}l@{}}CCMD-3: (Chinese psychiatric Association, 2001), \\ ICD-10\end{tabular} & Resting-state (eyes closed)                  \\
                     &                               &                                           & 26 (10F, 16M) healthy             & Healthy: 37.5                  &                                                                                                    &                                              \\
\rowcolor[HTML]{F2F2F2} 
\cite{20160801982396}             & Yes                           & Did not mention                           & 12 (7F, 5M) depressed             & Did not mention                & Did not mention                                                                                    & Resting-state                                \\
\rowcolor[HTML]{F2F2F2} 
                     &                               &                                           & 16 (7F, 9M) healthy               &                                &                                                                                                    &                                              \\
\cite{20070710426068}             & Yes                           & 5 Minutes                                 & 10 depressed                      & 20 to 83 with an average of 43 & Did not mention                                                                                    & Resting-state (eyes closed)                  \\
                     &                               &                                           & 11 healthy                        &                                &                                                                                                    &                                              \\
                     &                               &                                           & (13F, 9M)                         &                                &                                                                                                    &                                              \\
\rowcolor[HTML]{F2F2F2} 
{}\cite{8101151}{}              & Yes                           & Did not mention                           & 25 depressed                      & Did not mention                & Did not mention                                                                                    & Resting-state with audio stimulation         \\
\rowcolor[HTML]{F2F2F2} 
                     &                               &                                           & 25 healthy                        &                                &                                                                                                    &                                              \\
\cite{Pockberger1985}             & Yes                           & 12 minutes                                & 10 (10M) depressed                & Total: 20 to 35 years          & DSM-III                                                                                            & Resting-state (eyes closed and open)         \\
                     &                               &                                           & 10 (10M) healthy                  & Depressed: Mean: 26.1          &                                                                                                    &                                              \\
                     &                               &                                           &                                   & Healthy: Mean: 25.7            &                                                                                                    &                                             
\end{tabular}}
\end{sidewaystable*}

\begin{sidewaystable*}[htp]
\caption{\label{tab:Table4-1} Summary of the most recent literature based on type and number of electrodes, their position, data acquisition device, and sampling frequency, impedance and the type of filters used}
\resizebox{1\textwidth}{!}{
}
\end{sidewaystable*}

\begin{sidewaystable*}[htp]
\caption{\label{tab:Table4-12} Continued - summary of the most recent literature based on type and number of electrodes, their position, data acquisition device, and sampling frequency, impedance and the type of filters used}
\resizebox{1\textwidth}{!}{%
}
\end{sidewaystable*}

\begin{sidewaystable*}[htp]
\caption{\label{tab:Table4-13} Continued - summary of the most recent literature based on type and number of electrodes, their position, data acquisition device, and sampling frequency, impedance and the type of filters used}
\resizebox{1\textwidth}{!}{%
               & 1000 Hz                     & Under 20 K\si{\ohm}                        & 1-40 Hz bandpass filter                                                                                                              &  \\
\cite{Duan2021}        & 64 channels                            & N/A                                                                                                                                                                                                 & Brain Products                                                                                            & 100 Hz                      & Did not mentioned                  & Did not mentioned                                                                                                                    &

\end{tabular}}
\end{sidewaystable*}

\begin{sidewaystable*}[htp]
\caption{\label{tab:Table4-14} Continued - summary of the most recent literature based on type and number of electrodes, their position, data acquisition device, and sampling frequency, impedance and the type of filters used}
\resizebox{1\textwidth}{!}{
}
\end{sidewaystable*}

\begin{sidewaystable*}[htp]
\caption{\label{tab:Table4-111} Continued - summary of the most recent literature based on type and number of electrodes, their position, data acquisition device, and sampling frequency, impedance and the type of filters used }
\resizebox{1\textwidth}{!}{%
}
\end{sidewaystable*}

\begin{sidewaystable*}[htp]
\caption{\label{tab:Table4-122} Continued - summary of the most recent literature based on type and number of electrodes, their position, data acquisition device, and sampling frequency, impedance and the type of filters used }
\resizebox{1\textwidth}{!}{%
                                                                                                                                                                      & 500 Hz                                                                                                                  & Did not mention                                                                                                    & 1-42 Hz FIR bandpass filter                                                                                                                                                      \\

\end{tabular}}
\end{sidewaystable*}

\begin{sidewaystable*}[htp]
\caption{\label{tab:Table4-133} Continued - summary of the most recent literature based on type and number of electrodes, their position, data acquisition device, and sampling frequency, impedance and the type of filters used }
\resizebox{1\textwidth}{!}{
                                                                                                                                                                      & 500 Hz                                                                                              & Under 2 K\si{\ohm}                                                                                                 & 1 Hz high pass filter                                                                                                                                    \\

\end{tabular}}
\end{sidewaystable*}

\begin{sidewaystable*}[htp]
\caption{\label{tab:Table4-144} Continued - summary of the most recent literature based on type and number of electrodes, their position, data acquisition device, and sampling frequency, impedance and the type of filters used }
\resizebox{1\textwidth}{!}{
                                                                                                                                            & 400 Hz                                                                                              & Under 5 K\si{\ohm}                                                                                                  & 0.5–39 Hz bandpass filter                                                                                                                              \\

\end{tabular}}
\end{sidewaystable*}

\begin{sidewaystable*}[htp]
\caption{\label{tab:Table4-155} Continued - summary of the most recent literature based on type and number of electrodes, their position, data acquisition device, and sampling frequency, impedance and the type of filters used }
\resizebox{1\textwidth}{!}{\begin{tabular}{lllllllll}

\multicolumn{1}{c}{\textit{\textbf{\#}}}             & \multicolumn{1}{c}{\textit{\textbf{\begin{tabular}[c]{@{}c@{}}Number of electrodes/\\ channels\end{tabular}}}} & \multicolumn{1}{c}{\textit{\textbf{Sensor position}}}                                                                                             & \multicolumn{1}{c}{\textit{\textbf{Name of the device}}}                                                                 & \multicolumn{1}{c}{\textit{\textbf{\begin{tabular}[c]{@{}c@{}}Sampling \\ frequency\end{tabular}}}} & \multicolumn{1}{c}{\textit{\textbf{\begin{tabular}[c]{@{}c@{}}Electrodes Type/\\ Impedance\end{tabular}}}} & \multicolumn{1}{c}{\textit{\textbf{Filter}}}                                                                 \\
                                                     &                                                                                                                &                                                                                                                                                   &                                                                                                                          &                                                                                                     &                                                                                                            &                                                                                                              \\
\rowcolor[HTML]{F2F2F2} 
\multicolumn{1}{c}{\cellcolor[HTML]{F2F2F2}{}\cite{Hosseinifard2013}{}} & 19 electrodes                                                                                                  & \begin{tabular}[c]{@{}l@{}}Fz, Cz, Pz, Fp1, Fp2, F3, F4, F7, F8, \\ C3, C4, T3, T4, P3, P4, T5, T6, O1 and O2\end{tabular}                        & Did not mention                                                                                                          & 256 Hz                                                                                              & Did not mention                                                                                            & \begin{tabular}[c]{@{}l@{}}0.5 Hz high pass and \\ 70 Hz low pass filter, \\ 50 Hz notch filter\end{tabular} \\
\cite{Frantzidis2012}                                             & 57 electrodes                                                                                                  & 57 electrodes on scalp                                                                                                                            & \begin{tabular}[c]{@{}l@{}}57 electrodes with Nihon Kohden JE-207A \\ (NIHON KOHDEN EUROPE GmbH) EEG system\end{tabular} & 500 Hz                                                                                              & Under 2 K\si{\ohm}                                                                                                  & Did not mention                                                                                              \\
\rowcolor[HTML]{F2F2F2} 
\cite{Puthankattil2012ClassificationEntropy}                                            & Fp1-T3 (left half) and Fp2-T4 (right half)                                                                   & Fp1-T3 (left half), Fp2-T4 (right half)                                                                                                           & Did not mention                                                                                                          & 256 Hz                                                                                              & Did not mention                                                                                            & 50 Hz Notch filter                                                                                           \\
\cite{Ku2012}                                             & 19 channels                                                                                                    & \begin{tabular}[c]{@{}l@{}}Fp1, Fp2, F7, F3, Fz, F4, F8, T3, C3, Cz, \\ C4, T4, T5, P3, Pz, P4, T6, O1, and O2\end{tabular}                       & \begin{tabular}[c]{@{}l@{}}MindSet EEG Systems \\ (NP-Q 10/20, NeuroPulse-Systems, Inc.)\end{tabular}                    & 256 Hz                                                                                              & Did not mention                                                                                            & \begin{tabular}[c]{@{}l@{}}Median and band-stop filters to \\ remove baseline and 60 Hz noise\end{tabular}   \\
\rowcolor[HTML]{F2F2F2} 
{}\cite{12424821}{}                                             & 16 electrodes                                                                                                  & \begin{tabular}[c]{@{}l@{}}Fp1, Fp2, F3, F4, C3, C4, P3, P4, O1, O2, \\ F7, F8, T3, T4, T5, and T6\end{tabular}                                   & \begin{tabular}[c]{@{}l@{}}16 Channel EEG data acquisition device \\ (Sunray, LQWY-N, Guangzhou, China)\end{tabular}   & 100 Hz                                                                                              & Ag/AgCl electrodes                                                                                         & 0.5-30 Hz Bandpass filter                                                                                    \\
\cite{20102413002260}                                             & 19 electrodes                                                                                                  & \begin{tabular}[c]{@{}l@{}}19 Electrodes based on 10-20 system, \\ frontal, parietal, temporal, occipital\end{tabular}                            & \begin{tabular}[c]{@{}l@{}}Cadwell Easy II \\ (Cadwell Industries Inc., USA) EEG system\end{tabular}                   & 400 Hz                                                                                              & Did not mention                                                                                            & 0.5-48 Hz bandpass filter                                                                                    \\
\rowcolor[HTML]{F2F2F2} 
\cite{Suhhova2009}                                             & 9 electrodes                                                                                                   & \begin{tabular}[c]{@{}l@{}}9 electrodes based on the international \\ 10-20 system, frontal, parietal, temporal, \\ occipital\end{tabular}        & \begin{tabular}[c]{@{}l@{}}Cadwell Easy II \\ (Cadwell Industries Inc., USA) EEG system\end{tabular}                   & 400 Hz                                                                                              & Did not mention                                                                                            & 0.5-48 Hz bandpass filter                                                                                    \\
\cite{Hinrikus2009}                                            & 9 EEG and 2 EOG electrodes                                                                                     & \begin{tabular}[c]{@{}l@{}}Frontal-Fp1, Fp2; parietal-P5, P4; \\ temporal-T3, T4; occipital-O1, O2 \\ and the reference electrode Cz\end{tabular} & \begin{tabular}[c]{@{}l@{}}Cadwell Easy II \\ (Cadwell   Industries Inc., USA) EEG system\end{tabular}                   & 400 Hz                                                                                              & Under 5 K\si{\ohm}                                                                                                  & 0.5-48 Hz bandpass filter                                                                                    \\
\rowcolor[HTML]{F2F2F2} 
\cite{Lee2007}                                             & 8 electrodes                                                                                                   & \begin{tabular}[c]{@{}l@{}}F3 and F4 (frontal), C3 and C4 (central), \\ T3 and T4 (temporal) and on O1 and \\ O2 (occipital) scalp\end{tabular}   & \begin{tabular}[c]{@{}l@{}}EEG measurement device, LEX3208   \\ (LAXTHA Inc, Korea)\end{tabular}                         & 256Hz                                                                                               & \begin{tabular}[c]{@{}l@{}}Ag/AgCl electrodes, \\ Under 5 K\si{\ohm} \end{tabular}                                  & \begin{tabular}[c]{@{}l@{}}0.6 Hz high pass, \\ 46 Hz Low pass and \\ 60Hz notch filter\end{tabular}         \\
\cite{Li2007}                                             & 16 channels                                                                                                    & \begin{tabular}[c]{@{}l@{}}Fp1, Fp2, F3, F4, C3, C4, P3, P4, O1, \\ O2, F7, F8, T3, T4, T5 and T6\end{tabular}                                    & \begin{tabular}[c]{@{}l@{}}16 channel EEG acquisition system Model LQWY-N, \\ (Sunray, China)\end{tabular}               & 100 Hz                                                                                              & Did not mention                                                                                            & 0.5 Hz high pass filter                                                                                      \\
\rowcolor[HTML]{F2F2F2} 
\cite{20160801982396}                                              & 16 channels                                                                                                    & Did not mention                                                                                                                                   & Did not mention                                                                                                          & Did not mention                                                                                     & Did not mention                                                                                            & Did not mention                                                                                              \\
                                                     &                                                                                                                &                                                                                                                                                   &                                                                                                                          &                                                                                                     &                                                                                                            &                                                                                                              \\
\multirow{-2}{*}{\cite{20070710426068}}                           & \multirow{-2}{*}{4 channels}                                                                                   & \multirow{-2}{*}{Fp1, Fp2, left earlobe and right earlobe}                                                                                        & \multirow{-2}{*}{\begin{tabular}[c]{@{}l@{}}4 Electrodes with Biopac MP100A \\ (BIOPAC Systems Inc., UK)\end{tabular}}   & \multirow{-2}{*}{1 kHz}                                                                             & \multirow{-2}{*}{\begin{tabular}[c]{@{}l@{}}Ag/AgCl electrodes, \\ under 2K\si{\ohm} \end{tabular}}                 & \multirow{-2}{*}{Did not mention}                                                                            \\
\rowcolor[HTML]{F2F2F2} 
{}\cite{8101151}{}                                              & 15 electrodes                                                                                                  & \begin{tabular}[c]{@{}l@{}}Fp1, Fp2, F3, F4, C3, C4, (C3-T5)/2, Fig. 4), \\ (C4-T6)/2, P3, P4, O1, O2, Pz, Cz, and Fz\end{tabular}                & Did not mention                                                                                                          & 500 Hz                                                                                              & Ag/AgCl electrodes                                                                                         & Did not mention                                                                                              \\
\cite{Pockberger1985}                                            & 19 electrodes                                                                                                  & Did not mention                                                                                                                                   & Did not mention                                                                                                          & Did not mention                                                                                     & Did not mention                                                                                            & Did not mention                                                                                             
\end{tabular}}
\end{sidewaystable*}

\begin{table*}[htp]
\caption{\label{tab:Table5-1} Types of features used in most recent literature }
\resizebox{1\textwidth}{!}{
}
\end{table*}

\begin{table*}[htp]
\caption{\label{tab:Table5-3} Continued - types of features used in most recent literature }
\resizebox{1\textwidth}{!}{%
}                                                       \\
\rowcolor[HTML]{F2F2F2} 
\cellcolor[HTML]{F2F2F2}                            & Alpha power variability, Relative gamma power                                                                                                                                                                             & Detrended fluctuation analysis                                                                                                                                                                                       \\
\rowcolor[HTML]{F2F2F2} 
\multirow{-2}{*}{\cellcolor[HTML]{F2F2F2}\cite{Bachmann2018}}  & Spectral asymmetry index                                                                                                                                                                                                  & Higuchi fractal dimension, Lempel-ziv complexity                                                                                                                                                                     \\
                                                    &                                                                                                                                                                                                                           &                                                                                                                                                                                                                      \\
                                                    &                                                                                                                                                                                                                           &                                                                                                                                                                                                                      \\
\multirow{-3}{*}{\cite{Shen2017ACriterion}}                          & \multirow{-3}{*}{Centroid frequency, Mean frequency, Max frequency}                                                                                                                                                       & \multirow{-3}{*}{C0 complexity, Correlation dimension, Renyi entropy}                                                                                                                                                \\
\rowcolor[HTML]{F2F2F2} 
\cellcolor[HTML]{F2F2F2}                            & \cellcolor[HTML]{F2F2F2}                                                                                                                                                                                                  & \cellcolor[HTML]{F2F2F2}                                                                                                                                                                                             \\
\rowcolor[HTML]{F2F2F2} 
\cellcolor[HTML]{F2F2F2}                            & \cellcolor[HTML]{F2F2F2}                                                                                                                                                                                                  & \cellcolor[HTML]{F2F2F2}                                                                                                                                                                                             \\
\rowcolor[HTML]{F2F2F2} 
\multirow{-3}{*}{\cellcolor[HTML]{F2F2F2}\cite{Zhao2017}}  & \multirow{-3}{*}{\cellcolor[HTML]{F2F2F2}Center frequency, Mean frequency, Max frequency}                                                                                                                                 & \multirow{-3}{*}{\cellcolor[HTML]{F2F2F2}C0 complexity, Lempel-ziv complexity, Permutation cntropy}                                                                                                                  \\
                                                    & Mean, Total and relative Power of different bands                                                                                                                                                                         &                                                                                                                                                                                                                      \\
                                                    & Spectral asymmetry Index (SASI) values                                                                                                                                                                                    &                                                                                                                                                                                                                      \\
\multirow{-3}{*}{\cite{Ritu2017}}                         & \begin{tabular}[c]{@{}l@{}}Event-related synchronization (ERS), \\ Event-related desynchronization (ERD)\end{tabular}                                                                                                     & \multirow{-3}{*}{}                                                                                                                                                                                                   \\
\rowcolor[HTML]{F2F2F2} 
\cellcolor[HTML]{F2F2F2}                            & Mean, variance, skewness, kurtosis (statistical features)                                                                                                                                                                 & Correlation dimension, Higuchi fractal dimension                                                                                                                                                                     \\
\rowcolor[HTML]{F2F2F2} 
\cellcolor[HTML]{F2F2F2}                            & \begin{tabular}[c]{@{}l@{}}Delta, theta, alpha, beta and gamma frequency sub-bands \\ (frequency features)\end{tabular}                                                                                                   & Katz, Lempel-ziv complexity                                                                                                                                                                                          \\
\rowcolor[HTML]{F2F2F2} 
\multirow{-3}{*}{\cellcolor[HTML]{F2F2F2}\cite{17991812}}  & Coefficients of 10th order of the AR model, RQA                                                                                                                                                                           &                                                                                                                                                                                                                      \\
                                                    & Full band, Alpha, beta, theta and gamma absolute powers                                                                                                                                                                   & Power spectrum entropy                                                                                                                                                                                               \\
                                                    & Full band, Alpha, beta, theta and gamma relative powers                                                                                                                                                                   & Gamma, theta, alpha and beta power spectrum entropy                                                                                                                                                                  \\
                                                    & \begin{tabular}[c]{@{}l@{}}Alpha, beta, theta and gamma absolute and \\ relative center frequencies\end{tabular}                                                                                                          & \begin{tabular}[c]{@{}l@{}}Kolmogorov entropy, C0 complexity, Shannon entropy, \\ Correlation integrals\end{tabular}                                                                                                 \\
\multirow{-4}{*}{\cite{Cai2017NO2}}                          & Peak-to-peak, Variance, Hjorth activity, Kurtosis, Inclination                                                                                                                                                            &                                                                                                                                                                                                                      \\
\rowcolor[HTML]{F2F2F2} 
\cite{Liao2017}                                          & Theta, alpha, beta and gamma band   power                                                                                                                                                                                 & \begin{tabular}[c]{@{}l@{}}Grssberger and Procaccia (GP) algorithm-based fractal dimension \\ (GPFD) or correlation dimension\end{tabular}                                                                           \\
\rowcolor[HTML]{F2F2F2} 
\cellcolor[HTML]{F2F2F2}                            & \cellcolor[HTML]{F2F2F2}                                                                                                                                                                                                  & \cellcolor[HTML]{F2F2F2}                                                                                                                                                                                             \\

\multirow{-1}{*}{\cite{Puthankattil2017}}                          & \multirow{-1}{*}{\begin{tabular}[c]{@{}l@{}}Mean amplitude, mean duration, coefficient of variation of duration, \\ coefficient of variation of amplitude and slope (time domain)\end{tabular}}                           & \multirow{-1}{*}{Relative wavelet energy (RWE), Wavelet entropy (WE)}                                                                                                                                                \\

\end{tabular}}
\end{table*}

\begin{table*}[htp]
\caption{\label{tab:Table5-4} Continued - types of features used in most recent literature }
\resizebox{1\textwidth}{!}{\begin{tabular}{lll}

\textit{\textbf{\#}}                               & \textit{\textbf{Linear features}}                                                                                                                                                          & \textit{\textbf{Non-linear features}}                                                                                                                                                        \\
                                                   &                                                                                                                                                                                            &                                                                                                                                                                                              \\
\rowcolor[HTML]{F2F2F2} 
\cellcolor[HTML]{F2F2F2}                            & \begin{tabular}[c]{@{}l@{}}Absolute and relative centroid frequency, \\ Absolute and relative power\end{tabular}                                                                                                          & \cellcolor[HTML]{F2F2F2}                                                                                                                                                                                             \\
\rowcolor[HTML]{F2F2F2} 
\cellcolor[HTML]{F2F2F2}                            & Centroid frequency, Kurtosis, Slope                                                                                                                                                                                       & \cellcolor[HTML]{F2F2F2}                                                                                                                                                                                             \\
\rowcolor[HTML]{F2F2F2} 
\multirow{-3}{*}{\cellcolor[HTML]{F2F2F2}\cite{Cai2016PervasiveCollector}}  & \begin{tabular}[c]{@{}l@{}}Absolute and relative centroid frequency asymmetry, Absolute and \\ relative power asymmetry\end{tabular}                                                                                      & \multirow{-3}{*}{\cellcolor[HTML]{F2F2F2}\begin{tabular}[c]{@{}l@{}}C0 complexity, Correlation dimension, Kolmogorov entropy, \\ Power spectrum entropy, Shannon entropy\end{tabular}}                               \\
                                                    &                                                                                                                                                                                                                           &                                                                                                                                                                                                                      \\
\multirow{-2}{*}{\cite{Bachmann2017}}                          & \multirow{-2}{*}{Power spectral density, EEG signal powers}                                                                                                                                                               & \multirow{-2}{*}{}                                                                                                                                                                                                   \\
\rowcolor[HTML]{F2F2F2} 
\cellcolor[HTML]{F2F2F2}                            & Mean, variance, skewness, kurtosis (statistical features)                                                                                                                                                                 & \cellcolor[HTML]{F2F2F2}                                                                                                                                                                                             \\
\rowcolor[HTML]{F2F2F2} 
\cellcolor[HTML]{F2F2F2}                            & \begin{tabular}[c]{@{}l@{}}Curve length, number of peaks, \\ average nonlinear energy (Morphological)\end{tabular}                                                                                                        & \cellcolor[HTML]{F2F2F2}                                                                                                                                                                                             \\
\rowcolor[HTML]{F2F2F2} 
\cellcolor[HTML]{F2F2F2}                            & Band power and relative band power, Band power asymmetry                                                                                                                                                                  & \cellcolor[HTML]{F2F2F2}                                                                                                                                                                                             \\
\rowcolor[HTML]{F2F2F2} 
\multirow{-4}{*}{\cellcolor[HTML]{F2F2F2}{}\cite{Puk2016}{}}   & Hjorth, Spectral edge frequency, Zero crossing                                                                                                                                                                            & \multirow{-4}{*}{\cellcolor[HTML]{F2F2F2}Wavelet entropy (WE)}     
\\

\cite{Mumtaz2017}  & Alpha interhemispheric asymmetry, EEG spectral power                                                                                                                          &                                                                                                                                                                            \\
\rowcolor[HTML]{F2F2F2} 
\cite{XiaoweiLi2016EEG-basedClassifiers}                                          & Hjorth (Activity, mobility, and complexity)                                                                                                                                                & Approximate entropy, C0   complexity, Correlation dimension                                                                                                                                  \\
\rowcolor[HTML]{F2F2F2} 
                                                   & \begin{tabular}[c]{@{}l@{}}Power spectrum density, Max power spectrum density, Mean square, \\ Peak to peak amplitude, Sumpower, Variance\end{tabular}                                     & \begin{tabular}[c]{@{}l@{}}Kolmogorov entropy, Lyapunov exponent, Lempel-Ziv complexity, \\ Permutation entropy, Spectral, Singular-value deposition entropy\end{tabular}                    \\
\cite{Erguzel2016}                                           & \begin{tabular}[c]{@{}l@{}}Cordance values from absolute power, relative power,\\ maximum absolute, maximum relative and normalized absolute and\\  normalized relative values\end{tabular} &                                                                                                                                                                                              \\
\rowcolor[HTML]{F2F2F2} 
\cellcolor[HTML]{F2F2F2}                           & \cellcolor[HTML]{F2F2F2}                                                                                                                                                                   & \cellcolor[HTML]{F2F2F2}                                                                                                                                                                     \\
\rowcolor[HTML]{F2F2F2} 
\multirow{-2}{*}{\cellcolor[HTML]{F2F2F2}\cite{20160201780626}} & \multirow{-2}{*}{\cellcolor[HTML]{F2F2F2}}                                                                                                                                                 & \multirow{-2}{*}{\cellcolor[HTML]{F2F2F2}Higuchi fractal dimension (HFD), Katz fractal dimension (KFD)}                                                                                      \\
\cite{15605052}                                           &                                                                                                                                                                                            & \begin{tabular}[c]{@{}l@{}}Higuchi fractal dimension (HFD), Kolmogorov entropy, \\ Katz fractal dimension (KFD), Lempel-ziv complexity, \\ Shannon entropy\end{tabular}                      \\
\rowcolor[HTML]{F2F2F2} 
\cellcolor[HTML]{F2F2F2}                           & \cellcolor[HTML]{F2F2F2}                                                                                                                                                                   & \cellcolor[HTML]{F2F2F2}                                                                                                                                                                     \\
\rowcolor[HTML]{F2F2F2} 
\multirow{-2}{*}{\cellcolor[HTML]{F2F2F2}\cite{Kalev2015}} & \multirow{-2}{*}{\cellcolor[HTML]{F2F2F2}}                                                                                                                                                 & \multirow{-2}{*}{\cellcolor[HTML]{F2F2F2}Lempel-ziv complexity, Multiscale lempel-ziv complexity}                                                                                            \\
                                                   &                                                                                                                                                                                            &                                                                                                                                                                                              \\
\multirow{-2}{*}{\cite{15798207}}                         & \multirow{-2}{*}{\begin{tabular}[c]{@{}l@{}}Variance, Max power, Sum power\\ (12 linear features in total. Some of the features are not mentioned)\end{tabular}}                           & \multirow{-2}{*}{\begin{tabular}[c]{@{}l@{}}Co-complexity, Kolmogorov entropy, Lyapunov entropy \\ (8 nonlinear features in total. Some of the features are \\ not mentioned.)\end{tabular}} \\
\rowcolor[HTML]{F2F2F2} 
\cite{15670461}                                           & Power spectrum                                                                                                                                                                             &                                                                                                                                                                                              \\
{}\cite{Mumtaz2015ADisorder}{}                                            & \begin{tabular}[c]{@{}l@{}}Inter-hemispheric asymmetries, Power of different frequency \\ bands of EEG signals\end{tabular}                                                                &                                                                                                                                                                                              \\
\rowcolor[HTML]{F2F2F2} 
\cellcolor[HTML]{F2F2F2}                           & \cellcolor[HTML]{F2F2F2}                                                                                                                                                                   & \cellcolor[HTML]{F2F2F2}                                                                                                                                                                     \\
\rowcolor[HTML]{F2F2F2} 
\multirow{-2}{*}{\cellcolor[HTML]{F2F2F2}\cite{Sun2015}} & \multirow{-2}{*}{\cellcolor[HTML]{F2F2F2}Gravity frequency, Relative power}                                                                                                                & \multirow{-2}{*}{\cellcolor[HTML]{F2F2F2}}                                                                                                                                                   \\
{}\cite{Bachmann2015}{}                                            &                                                                                                                                                                                            & Lempel-ziv complexity                                                                                                                                                                        \\
\rowcolor[HTML]{F2F2F2} 
\cellcolor[HTML]{F2F2F2}                           & \cellcolor[HTML]{F2F2F2}                                                                                                                                                                   & \cellcolor[HTML]{F2F2F2}                                                                                                                                                                     \\
\rowcolor[HTML]{F2F2F2} 
\multirow{-2}{*}{\cellcolor[HTML]{F2F2F2}\cite{Katyal2014}} & \multirow{-2}{*}{\cellcolor[HTML]{F2F2F2}Power spectrum, Mean, standard deviation and median}                                                                                              & \multirow{-2}{*}{\cellcolor[HTML]{F2F2F2}Entropy}                                                                                                                                            \\
                                                   &                                                                                                                                                                                            &                                                                                                                                                                                              \\
\multirow{-2}{*}{\cite{Faust2014}}                         & \multirow{-2}{*}{Wavelet packet decomposition}                                                                                                                                             & \multirow{-2}{*}{\begin{tabular}[c]{@{}l@{}}Bispectral phase entropy, Renyi entropy, Sample entropy, \\ Approximate entropy\end{tabular}}                                                    \\
\rowcolor[HTML]{F2F2F2} 
\cellcolor[HTML]{F2F2F2}                           & \cellcolor[HTML]{F2F2F2}                                                                                                                                                                   & \cellcolor[HTML]{F2F2F2}                                                                                                                                                                     \\
\rowcolor[HTML]{F2F2F2} 
\multirow{-2}{*}{\cellcolor[HTML]{F2F2F2}\cite{13386381}} & \multirow{-2}{*}{\cellcolor[HTML]{F2F2F2}Absolute power, Relative power, Center frequency}                                                                                                 & \multirow{-2}{*}{\cellcolor[HTML]{F2F2F2}\begin{tabular}[c]{@{}l@{}}C0 complexity, Correlation dimension, Lyapunov exponent, \\ Lempel-ziv complexity\end{tabular}}                          \\
                                                   &                                                                                                                                                                                            &                                                                                                                                                                                              \\
\multirow{-2}{*}{\cite{Suhhova2013}}                         & \multirow{-2}{*}{Power spectral density, EEG signal powers}                                                                                                                                & \multirow{-2}{*}{}                                                                                                                                                                           \\
\rowcolor[HTML]{F2F2F2} 
\cellcolor[HTML]{F2F2F2}                           & \cellcolor[HTML]{F2F2F2}                                                                                                                                                                   & Correlation dimension, Detrended fluctuation analysis                                                                                                                                        \\
\rowcolor[HTML]{F2F2F2} 
\multirow{-2}{*}{\cellcolor[HTML]{F2F2F2}{}\cite{Hosseinifard2013}{}} & \multirow{-2}{*}{\cellcolor[HTML]{F2F2F2}EEG band powers}                                                                                                                                  & Higuchi fractal dimension, Lyapunov exponent                                                                                                                                                 \\
\cite{Frantzidis2012}                                           &                                                                                                                                                                                            & Relative wavelet entropy (RWE)                                                                                                                                                               \\
\rowcolor[HTML]{F2F2F2} 
\cite{Puthankattil2012ClassificationEntropy}                                           &                                                                                                                                                                                            & Relative wavelet energy (RWE)                                                                                                                                                                \\
\cite{Ku2012}                                           & Two-dimensional Fourier transform                                                                                                                                                          & Approximate entropy                                                                                                                                                                          \\
\rowcolor[HTML]{F2F2F2} 
\cite{20102413002260}                                            & Power spectral density (PSD), EEG power, Spectral asymmetry (SA)                                                                                                                           &                                                                                                                                                                                              \\
\cite{Hinrikus2009}                                           & \begin{tabular}[c]{@{}l@{}}Spectral Asymmetry index, Inter-hemispheric asymmetry, \\ Inter-hemispheric coherence\end{tabular}                                                              &                                                                                                                                                                                              \\
\rowcolor[HTML]{F2F2F2} 
\cite{Li2007}                                          &                                                                                                                                                                                            & Wavelet entropy (WE)                                                                                                                                                                         \\
\cite{20070710426068}                                           & Frontal Brain Asymmetry (FBA) ratios                                                                                                                                                       &                                                                                                                                                                                              \\
\rowcolor[HTML]{F2F2F2} 
\cite{Pockberger1985}                                           & Averaged power                                                                                                                                                                             &

\end{tabular}}
\end{table*}

\pagebreak
\section*{Acknowledgments}
This work was supported in part by the UK Dementia Research Institute through UK DRI Ltd under the UKDRI-7204 award.  
\section*{Conflict of interest}
{The authors declare no conflict of interest.


\bibliography{ref}

\bibliographystyle{IEEEtran}

\newpage
\section*{Author Biographies}

\begin{wrapfigure}{L}{1in}
    \vspace{-10pt}
    \includegraphics[width=1.6in,height=1.6in,clip,keepaspectratio]{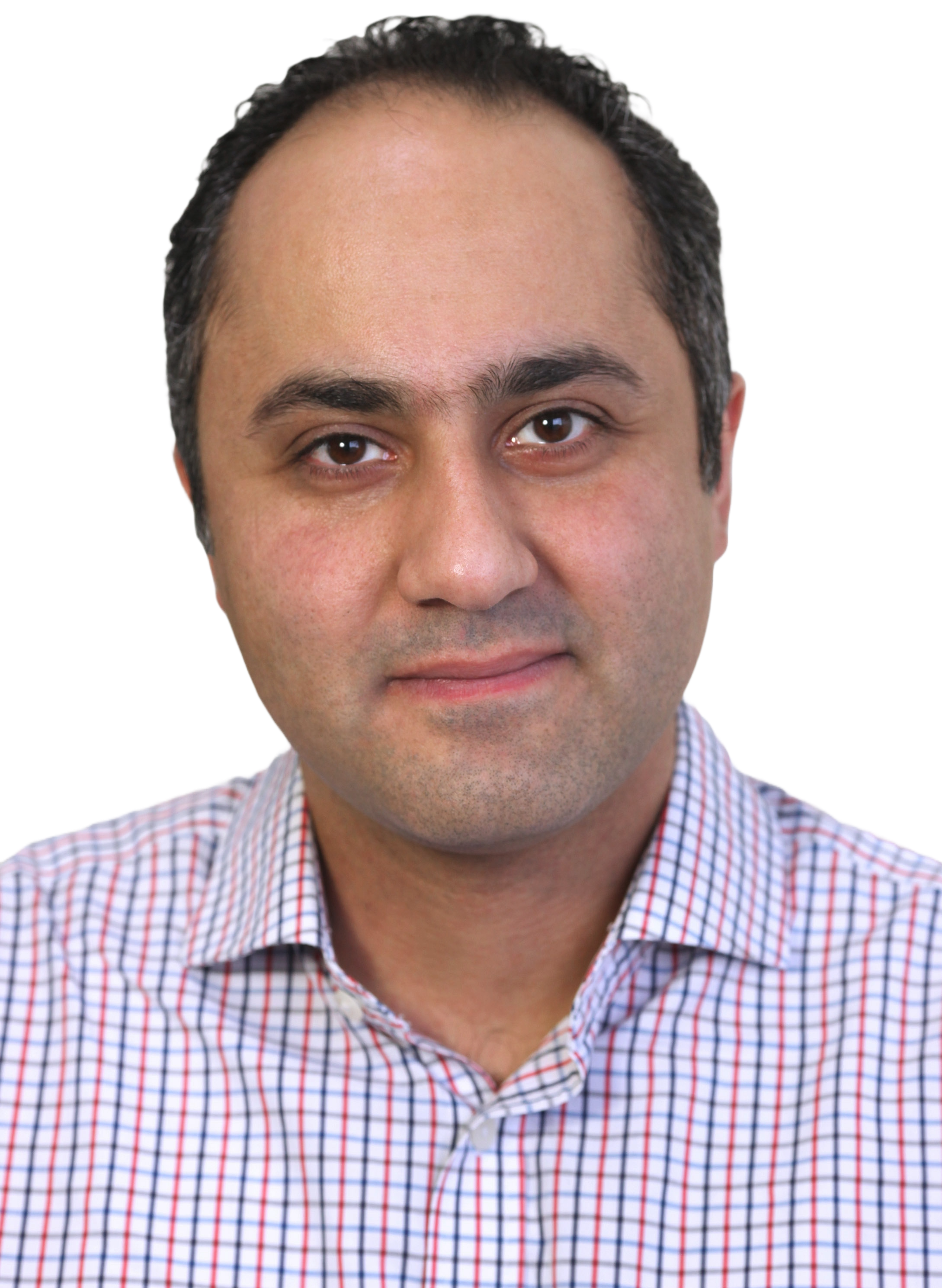}
    \vspace{-5pt}
\end{wrapfigure}
\textbf{Amir Nassibi} is currently a Postdoctoral Research Associate at the Electrical and Electronic Engineering Department, Imperial College London. He received his Ph.D. from Imperial College London and an MSc in Advanced Digital Systems from the University of Hertfordshire. His research interests include biosignal processing, machine learning algorithms, wearable devices, and embedded system design. He is interested in developing a career in research while maintaining his interest in mental health diagnosis using machine intelligence. His research has been supported by the UK Dementia Research Institute through UK DRI Ltd under the UKDRI-7204 award.\\

\begin{wrapfigure}{L}{1in}
    \vspace{-10pt}
    \includegraphics[width=1.35in,height=1.35in,clip,keepaspectratio]{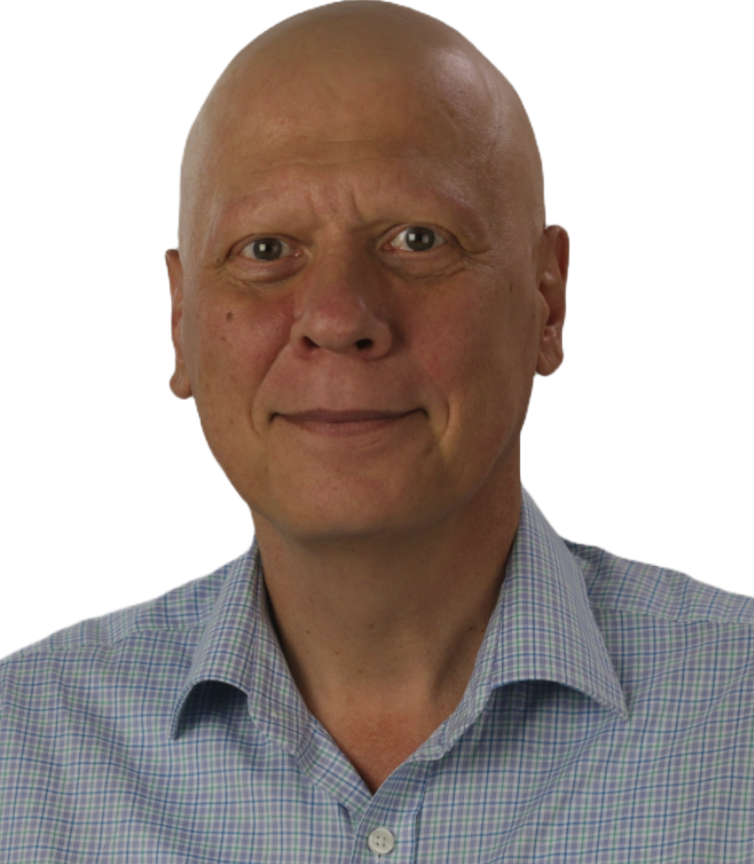}
    \vspace{-5pt}
\end{wrapfigure}
\textbf{Christos Papavassiliou} received the B.Sc. degree in physics from the Massachusetts Institute of Technology and the Ph.D. degree in applied physics from Yale University. He is currently with the Electrical Engineering Department, Imperial College London. His research focuses on memristor applications, sensor devices and systems, and antenna array technology. He has contributed to over 70 publications on weak localization, GaAs MMICs, and RFIC.\\

\begin{wrapfigure}{L}{1in}
    \vspace{-10pt}
    \includegraphics[width=1.5in,height=1.5in,clip,keepaspectratio]{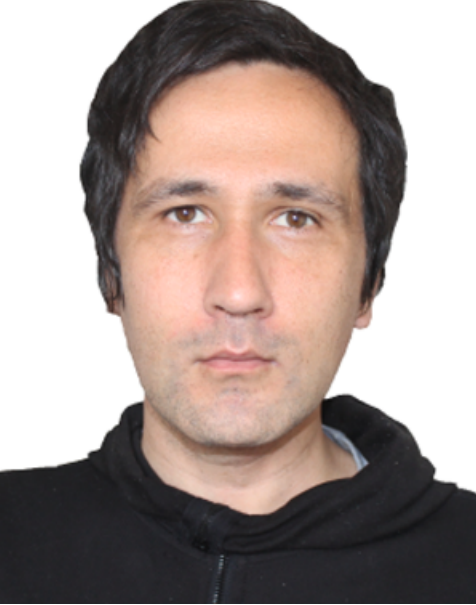}
    \vspace{-5pt}
\end{wrapfigure}
\textbf{Ildar Rakhmatulin} received a Ph.D. degree in Electrical Engineering from South Ural State University, Russia, in 2015. Since 2024, he has been a Research Associate in Neurotechnology Systems at the School of Engineering, University of Edinburgh. His research interests include brain-computer interfaces, signal processing, and machine learning.\\

\newpage

\begin{wrapfigure}{L}{1in}
    \vspace{-10pt}
    \includegraphics[width=1.65in,height=1.65in,clip,keepaspectratio]{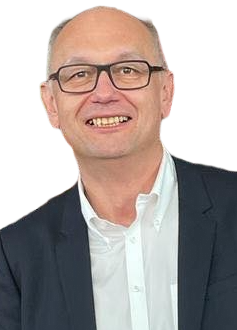}
    \vspace{-5pt}
\end{wrapfigure}
\textbf{Danilo Mandic} (Fellow, IEEE) received the Ph.D. degree in nonlinear adaptive signal processing from Imperial College London in 1999. He is currently a Professor of Machine Intelligence at Imperial College London. His publication record includes two research monographs: *Recurrent Neural Networks for Prediction: Learning Algorithms, Architectures and Stability* (Wiley, 2001) and *Complex Valued Nonlinear Adaptive Filters: Noncircularity, Widely Linear and Neural Models* (Wiley, 2009). He has also edited *Signal Processing Techniques for Knowledge Extraction and Information Fusion* (Springer, 2008) and a two-volume research monograph *Tensor Networks for Dimensionality Reduction and Large Scale Optimization* (Now Publishers, 2016 and 2017). He received the Denis Gabor Award for Outstanding Achievements in Neural Engineering from the International Neural Networks Society and the 2018 Best Paper Award from IEEE Signal Processing Magazine for his article on tensor decompositions. He has served as an Associate Editor for IEEE Transactions on Circuits and Systems—II: Express Briefs, IEEE Transactions on Signal Processing, IEEE Transactions on Neural Networks and Learning Systems, and IEEE Transactions on Signal and Information Processing Over Networks.\\

\begin{wrapfigure}{L}{1in}
    \vspace{-10pt}
    \includegraphics[width=1.45in,height=1.45in,clip,keepaspectratio]{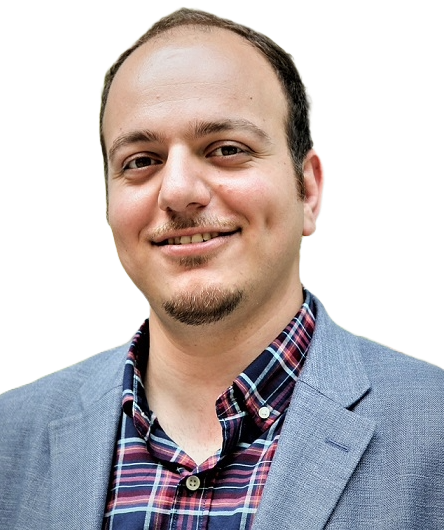}
    \vspace{-5pt}
\end{wrapfigure}
\textbf{S. Farokh Atashzar} is an Assistant Professor at New York University, jointly appointed at the Department of Electrical and Computer Engineering and the Department of Mechanical and Aerospace Engineering. He is also affiliated with NYU WIRELESS and the NYU Center for Urban Science and Progress (CUSP). At NYU, he leads the *Medical Robotics and Interactive Intelligent Technologies Lab*. Prior to joining NYU, he was a postdoctoral scientist at Imperial College London and, before that, a postdoctoral associate at the Canadian Surgical Technologies and Advanced Robotics (CSTAR) Center in Canada. He was the recipient of the NSERC Postdoctoral Fellowship Award and was ranked fifth nationally in Canada in the Electrical and Computer Engineering Sector. His research focus includes rehabilitation robotics, biosignal processing, and human-machine interfaces. He has received several awards, including the 2021 Outstanding Associate Editor Award for IEEE Robotics and Automation Letters. His research has been supported by the US National Science Foundation.

\end{document}